\documentclass[aps,prd,reprint,onecolumn,superscriptaddress,nofootinbib,floatfix,10pt]{revtex4-2}

\usepackage{amsmath,amssymb,bm}
\usepackage{graphicx}
\usepackage{float}
\usepackage{booktabs}
\usepackage{xcolor}
\usepackage[colorlinks=true,allcolors=blue]{hyperref}

\graphicspath{{figures/}}

\newcommand{\muB}{\mu_B}
\newcommand{\snn}{\sqrt{s_{NN}}}

\begin{document}

\title{High-order temperature fluctuations in hot and dense QCD matter from multi-charge holography}

\author{Yidian Chen}
\email{chenyidian@hznu.edu.cn}
\affiliation{School of Physics, Hangzhou Normal University, Hangzhou, 311121, China}
\author{Peiyu Yan}
\email{2026111030031@stu.hznu.edu.cn}
\affiliation{School of Physics, Hangzhou Normal University, Hangzhou, 311121, China}
\author{Zhibin Li}
\email{lizhibin@zzu.edu.cn}
\thanks{Corresponding author}
\affiliation{Institute for Astrophysics, School of Physics, Zhengzhou University, Zhengzhou 450001, China}

\begin{abstract}
We investigate temperature cumulants from second to sixth order across the QCD phase diagram using a lattice-calibrated multi-charge Einstein--Maxwells--dilaton model under heavy-ion charge constraints.
Near the model's critical end point at $(\mu_B,T)\simeq(547,112)$~MeV, the enhanced heat capacity produces a local suppression of the temperature variance.
The higher-order cumulants develop sign reversals on the high-temperature side of the transition, with the onset moving to lower baryon chemical potentials as the order increases.
Ratios of different cumulant orders provide a consistency check of the expected mean-field critical behavior.
At fixed chemical potentials, their limits on both sides of the critical point agree with the predictions for the critical-isotherm exponent $\delta=3$.
The freeze-out analysis reveals nonmonotonic collision-energy dependence that need not be accompanied by sign reversals.
On a representative trajectory, the variance reaches a local minimum near $\snn\simeq6$~GeV, with its position and depth depending on the freeze-out path.
Temperature cumulants probe the thermal response and its critical scaling, providing information complementary to conserved-charge fluctuations for exploring the QCD phase diagram.
\end{abstract}

\maketitle

\section{Introduction}
\label{sec:intro}

Quantum chromodynamics (QCD) predicts that strongly interacting matter undergoes a transition from hadronic matter to a quark-gluon plasma at high temperatures.
At vanishing baryon chemical potential and physical quark masses, lattice QCD has established that this transition is an analytic crossover~\cite{Aoki:2006we,Borsanyi:2013bia,HotQCD:2014kol}.
At larger $\muB$, effective-model, functional, and holographic studies suggest a first-order transition line terminating at a critical end point (CEP)~\cite{Stephanov:1998dy,Stephanov:1999zu,Qin:2010nq,Fu:2019hdw,Gao:2020fbl,Critelli:2017ouy,Cai:2022omk,Li:2026lxx}, although its existence and location remain open questions.
Recent lattice analyses have constrained the possible CEP region through entropy-contour continuation combined with chemical freeze-out information~\cite{Borsanyi:2025dyp}.
Complementary estimates have been obtained by extrapolating Lee--Yang edge singularities on finite lattices~\cite{Clarke:2024ugt}.
Relativistic heavy-ion collisions provide experimental access to this region of the QCD phase diagram, with decreasing collision energy generally corresponding to larger $\muB$ at chemical freeze-out~\cite{Adamczyk:2017iwn}.
Establishing whether the CEP exists and determining its location are major goals of the beam energy scan program at the Relativistic Heavy Ion Collider~\cite{Luo:2017faz,STAR:2020tga,STAR:2025zdq}.
The search for the CEP also motivates experimental programs at HIAF~\cite{Zhou:2022pxl}, FAIR~\cite{CBM:2016kpk}, and NICA~\cite{MPD:2022qhn}.

A major strategy in the experimental search for the CEP is to measure event-by-event fluctuations of conserved charges, particularly net-proton cumulants as proxies for net-baryon fluctuations~\cite{Luo:2017faz,Bzdak:2019pkr,STAR:2020tga,STAR:2021fge,STAR:2025zdq}.
In equilibrium, conserved-charge cumulants characterize the response of the pressure to chemical potentials.
Holographic studies, including our earlier work, have illustrated how critical enhancement of conserved-charge susceptibility ratios can produce peaks along suitable freeze-out trajectories~\cite{Li:2017ple,Li:2026lxx}.
Temperature fluctuations probe the complementary thermal response, with a variance inversely related to the heat capacity~\cite{Stodolsky:1995ds,Shuryak:1997yj,Stephanov:1999zu}.
Event-by-event fluctuations of the mean transverse momentum offer a possible experimental connection~\cite{Gavin:2003cb,Gardim:2019xjs,Giacalone:2020lbm}.
Recent measurements have extended this program to higher moments and the separation of geometrical and intrinsic contributions at the LHC~\cite{ALICE:2023tej,ATLAS:2024jvf}, while STAR has reported a nonmonotonic collision-energy dependence of two-particle transverse-momentum correlations at RHIC~\cite{STAR:2026vjv}.
A quantitative relation to equilibrium temperature cumulants requires dynamical modeling of initial-state geometry, collective expansion, volume fluctuations, hadronic interactions, and experimental acceptance~\cite{Reichert:2026ztj,Liu:2026lnr}.

A framework that generates temperature cumulants of arbitrary order from entropy derivatives of a Legendre-transformed thermodynamic potential has recently been developed~\cite{Chen:2025vwl}.
In this construction, temperature is the variable conjugate to entropy, and the derivatives are taken at fixed volume and chemical potentials.
Its application to a $2+1$-flavor low-energy effective theory within the functional renormalization group (fRG) revealed a pronounced suppression of temperature fluctuations from the hadronic to the quark-gluon-plasma regime, accompanied by negative skewness, with calculations extending to sixth-order cumulants and normalized cumulant ratios.
Subsequent studies in Polyakov-loop extended Nambu--Jona-Lasinio (PNJL) models found nonmonotonic higher-order ratios across the chiral transition and a local minimum of the variance along freeze-out trajectories near the critical region~\cite{Liu:2025cpj,Liu:2026pro}.
Calculations in ideal and interacting hadron resonance gases have further provided hadronic baselines and identified variance suppression and a skewness sign change along the Widom line of the nuclear liquid--gas transition~\cite{Mallick:2026dsc}.

Holographic QCD has been developed in several bottom-up frameworks, including improved holographic QCD (IHQCD)~\cite{Gursoy:2007cb,Gursoy:2007er,Gursoy:2009jd}, V-QCD~\cite{Jarvinen:2011qe,Alho:2013hsa}, and dynamical holographic QCD~\cite{Li:2013oda,Chelabi:2015cwn}.
Holographic Einstein--Maxwell--dilaton (EMD) models provide a framework for constructing thermodynamically consistent equations of state for strongly coupled QCD matter~\cite{Gubser:2008ny,DeWolfe:2010he,Cai:2012xh,Rougemont:2015wca,Grefa:2021qvt,Chen:2024ckb} and exploring its phase structure at finite temperature and density~\cite{Knaute:2017opk,Critelli:2017ouy,Cai:2022omk,Hippert:2023bel,Zhao:2023gur,Zhu:2025gxo}.
For temperature cumulants, a useful feature is that the entropy density is obtained directly from the black-hole horizon area, reducing the required order of numerical differentiation by one compared with a calculation based on the pressure.
We employ the multi-charge Einstein--Maxwells--dilaton (EMsD) model developed in Ref.~\cite{Li:2026lxx}, which extends the refined EMD construction of Ref.~\cite{Li:2025lmp} by introducing three flavor-resolved gauge fields.
Calibrated against zero-density lattice data, this model describes the coupled conserved charges $(B,Q,S)$ and allows us to impose strangeness neutrality and the nuclear charge-to-baryon ratio, including the associated nonzero $\mu_Q$ and $\mu_S$.
The expected mean-field behavior of the CEP in this classical-gravity setup provides a benchmark for testing the critical scaling of the temperature-cumulant hierarchy~\cite{DeWolfe:2010he,Chen:2018vty}.

In this work, we investigate temperature cumulants through sixth order across the phase diagram of the lattice-calibrated multi-charge model under heavy-ion charge constraints.
We relate variance suppression and higher-order sign changes to the thermal response of the medium and examine how different chemical freeze-out trajectories sample these structures.
We further connect the cumulant hierarchy to local critical scaling near the CEP and test this connection using ratios of cumulants.

The paper is organized as follows.
Sec.~\ref{sec:model} summarizes the holographic model, its lattice calibration, and the phase diagram under heavy-ion charge constraints.
Sec.~\ref{sec:cumulants} defines the temperature cumulants, derives their general thermodynamic properties, and describes the numerical methods.
Sec.~\ref{sec:results} examines $c_2$--$c_4$ at fixed chemical potentials, over the $T$--$\muB$ plane, and along chemical freeze-out trajectories, relating their characteristic structures to the thermal response of the medium.
Sec.~\ref{sec:higher} extends the analysis to $c_5$, $c_6$, and the normalized ratios $c_n/c_2^{\,n-1}$, and investigates the critical scaling of the cumulant hierarchy and the extraction of the critical-isotherm exponent $\delta$ from temperature cumulants.
Sec.~\ref{sec:conclusions} presents our discussion and conclusions.

\section{The multi-charge holographic model}
\label{sec:model}

\subsection{Action and black-brane thermodynamics}

We use the EMsD model of Ref.~\cite{Li:2026lxx}, based on the refined EMD construction of Ref.~\cite{Li:2025lmp}.
The bulk dilaton describes the nonconformal dynamics of the medium, while three Abelian gauge fields source the conserved flavor currents. 
In the Einstein frame, the five-dimensional action is
\begin{equation}
 S_{\mathrm{EMsD}}=\frac{1}{2\kappa_5^2}\int d^5x\sqrt{-g}
 \left[R-\frac12(\partial\phi)^2-V(\phi)
 -\frac14\sum_{q=u,d,s}Z_q(\phi)F_{q\,MN}F_q^{MN}\right],
 \label{eq:action}
\end{equation}
where $F_q=dA_q$, $\phi$ is the dilaton, and $\kappa_5^2$ is the five-dimensional gravitational coupling. 
We set the asymptotic AdS radius to unity.
The temporal boundary values of $A_q$ specify the flavor chemical potentials $\mu_q$.
Their linear combinations give the baryon, electric-charge, and strangeness chemical potentials $(\mu_B,\mu_Q,\mu_S)$, allowing the model to implement strangeness neutrality and the nuclear charge-to-baryon ratio relevant to heavy-ion collisions.

The dilaton potential and gauge-kinetic functions take the forms
\begin{align}
 V(\phi)&=-12\cosh(v_1\phi)
       +\left(6v_1^2-\frac32\right)\phi^2+v_2\phi^6+v_3\phi^8,
 \label{eq:model-potential}\\
 Z_q(\phi)&=\left[\cosh\!\left(c_{q1}\phi^{c_{q2}}\right)
                    +c_{q3}\phi^2\right]^{-1},
 \label{eq:model-couplings}
\end{align}
where $v_1$, $v_2$, and $v_3$ are the potential parameters, and $c_{qi}$ ($i=1,2,3$) control the coupling of each flavor gauge field to the dilaton.
The light flavors share the same couplings, $Z_u=Z_d$, while the strange sector has a separate parameter set. 
The potential behaves as $V(\phi)=-12-3\phi^2/2+O(\phi^4)$ near the boundary, corresponding to $m^2=-3$ and a boundary operator of dimension three in the standard quantization. 
The dilaton source $\phi_s$ fixes the physical energy scale, and the finite
counterterm coefficient $b_{\mathrm{ct}}$ specifies the renormalization prescription for the pressure and energy density. 
All model parameters are collected in Tab.~\ref{tab:model-parameters}.

\begin{table}[tb]
\centering
\caption{Parameters of the EMsD model adopted from Refs.~\cite{Li:2025lmp,Li:2026lxx}.
The potential parameters, gravitational coupling, dilaton source, and finite counterterm coefficient are common to all flavors; the gauge-kinetic parameters are specified separately for the light and strange sectors. The AdS radius is set to unity.}
\label{tab:model-parameters}
\begin{tabular}{lcc}
\toprule
Parameter & Light flavors ($u,d$) & Strange flavor ($s$) \\
\midrule
$v_1$ & \multicolumn{2}{c}{0.71} \\
$v_2$ & \multicolumn{2}{c}{$3.7\times10^{-3}$} \\
$v_3$ & \multicolumn{2}{c}{$2.8\times10^{-5}$} \\
$\kappa_5^2$ & \multicolumn{2}{c}{$3.36\pi$} \\
$\phi_s$ [MeV] & \multicolumn{2}{c}{1085} \\
$b_{\mathrm{ct}}$ & \multicolumn{2}{c}{$-0.27435$} \\
\midrule
$c_{q1}$ & 0.042 & 0.34 \\
$c_{q2}$ & 3.7 & 1.97 \\
$c_{q3}$ & $-0.04$ & $-0.06$ \\
\bottomrule
\end{tabular}
\end{table}
With this parameter set, the model describes the zero-density lattice equation of state and conserved-charge susceptibilities~\cite{Li:2026lxx}.

For homogeneous equilibrium states, the ultraviolet divergences of the on-shell action are removed by adding boundary terms at a radial cutoff $r=r_c$~\cite{deHaro:2000vlm}.
Following the holographic renormalization prescription of Refs.~\cite{Li:2025lmp,Li:2026lxx}, we define
\begin{align}
 S_{\mathrm{ren}}&=\lim_{r_c\to\infty}
 \left[S_{\mathrm{EMsD}}^{(r_c)}+S_{\partial}^{(r_c)}\right],
 \label{eq:model-renormalized-action}\\
 S_{\partial}^{(r_c)}&=\frac{1}{2\kappa_5^2}\int_{r=r_c}d^4x\sqrt{-h}
 \left[2K-6-\frac12\phi^2-b_{\mathrm{ct}}\phi^4
 +\frac{2+V^{(4)}(0)}{24}\phi^4\ln r_c\right].
 \label{eq:model-boundary-action}
\end{align}
Here $h_{\mu\nu}$ is the induced metric on the cutoff surface, and $K$ is the trace of its extrinsic curvature defined with the outward-pointing unit normal. 
The $2K$ contribution is the Gibbons--Hawking boundary term. The constant, quadratic, and logarithmic counterterms cancel the ultraviolet divergences, while $-b_{\mathrm{ct}}\phi^4$ is a finite counterterm fixing the vacuum normalization of the thermodynamic potential. 
For the even potential in Eq.~\eqref{eq:model-potential}, $V^{(3)}(0)=0$ and $V^{(4)}(0)=-12v_1^4$, so the cubic counterterm is absent. 
The value of $b_{\mathrm{ct}}$ is the one adopted in Refs.~\cite{Li:2025lmp,Li:2026lxx} and listed in Tab.~\ref{tab:model-parameters}.

The homogeneous black-brane solutions are described by the ansatz
\begin{equation}
 ds^2=-e^{-\eta(r)}f(r)\,dt^2+\frac{dr^2}{f(r)}+r^2d\bm x^2,
 \qquad \phi=\phi(r),\qquad A_q=A_{qt}(r)\,dt .
 \label{eq:metric}
\end{equation}
The conformal boundary lies at $r\to\infty$, and the outer horizon is at $f(r_h)=0$.
In this gauge the independent radial field equations can be written as
\begin{equation}
\begin{aligned}
 &\eta'+\frac{r}{3}\phi'^2=0,\\
 &\partial_r\!\left[r^3e^{\eta/2}Z_q A_{qt}'\right]=0,\\
 &\phi''+\left(\frac{3}{r}+\frac{f'}{f}-\frac{\eta'}{2}\right)\phi'
 -\frac{V_{,\phi}}{f}
 +\frac{e^\eta}{2f}\sum_q Z_{q,\phi}(A_{qt}')^2=0,\\
 &f'+\left(\frac{2}{r}-\frac{\eta'}{2}\right)f
 +\frac{r}{3}V+\frac{re^\eta}{6}\sum_q Z_q(A_{qt}')^2=0.
\end{aligned}
 \label{eq:model-eom}
\end{equation}
Here primes denote radial derivatives, and a subscript $\phi$ denotes differentiation with respect to the dilaton. 
The Maxwell equations identify three conserved radial electric fluxes,
\begin{equation}
 \mathcal Q_q=r^3e^{\eta/2}Z_q(\phi)A_{qt}'=2\kappa_5^2\rho_q,
 \label{eq:model-flux}
\end{equation}
where $\rho_q$ is the boundary flavor-number density.

Regularity requires finite scalar and metric data at the horizon, and we choose the regular gauge $A_{qt}(r_h)=0$. 
At the boundary, the conditions $f/r^2\to1$ and $\eta\to0$ fix the asymptotic geometry and the normalization of boundary time.
The near-boundary matter-field expansions are
\begin{equation}
\begin{aligned}
 \phi(r)&=\frac{\phi_s}{r}+\frac{\phi_v+\gamma\,\phi_s^3\ln r}{r^3}+o(r^{-3}),\\
 A_{qt}(r)&=\mu_q-\frac{\kappa_5^2\rho_q}{r^2}+o(r^{-2}).
\end{aligned}
 \label{eq:model-uv}
\end{equation}
Here $\phi_v$ is the normalizable scalar coefficient and $\mu_q$ is the potential difference between the horizon and the boundary.
The coefficient $\gamma$ of the logarithmic term is fixed by the near-boundary field equations and is responsible for the logarithmic counterterm in Eq.~\eqref{eq:model-boundary-action},
With boundary time normalized as above, the temperature and entropy density are
\begin{equation}
 T=\frac{e^{-\eta(r_h)/2}|f'(r_h)|}{4\pi},\qquad
 s=\frac{2\pi}{\kappa_5^2}r_h^3 .
 \label{eq:TS-horizon}
\end{equation}
The entropy is obtained algebraically from the horizon area.

The renormalized on-shell action and boundary stress tensor yield the pressure $p$ and energy density $\varepsilon$, which satisfy the following thermodynamic identities at fixed dilaton source:
\begin{equation}
 dp=s\,dT+\sum_q\rho_q\,d\mu_q,\qquad
 \varepsilon+p=Ts+\sum_q\mu_q\rho_q .
 \label{eq:model-first-law}
\end{equation}
At fixed $T$ and $\bm\mu=(\mu_B,\mu_Q,\mu_S)$, equilibrium selects the black-brane branch with the largest renormalized pressure.

\subsection{Conserved charges and phase structure}

To express these thermodynamic relations in terms of the charges relevant to heavy-ion collisions, we change from the flavor basis $(u,d,s)$ to the baryon-number, electric-charge, and strangeness basis $(B,Q,S)$. 
The chemical potentials and charge densities in the $(B,Q,S)$ basis are
\begin{align}
 \mu_B&=\mu_u+2\mu_d,& \mu_Q&=\mu_u-\mu_d,&
 \mu_S&=\mu_d-\mu_s,\nonumber\\
 \rho_B&=\frac{\rho_u+\rho_d+\rho_s}{3},&
 \rho_Q&=\frac{2\rho_u-\rho_d-\rho_s}{3},&
 \rho_S&=-\rho_s .
 \label{eq:charge-map}
\end{align}
These transformations preserve $\sum_q\mu_q\rho_q=\sum_{i=B,Q,S}\mu_i\rho_i$ and the corresponding differential form, so the thermodynamic identities in Eq.~\eqref{eq:model-first-law} retain the same form in the $(B,Q,S)$ basis.

To describe the charge composition relevant to heavy-ion collisions, we impose strangeness neutrality and a charge-to-baryon ratio of $0.4$~\cite{Bazavov:2012vg},
\begin{equation}
 \rho_S=0,\qquad \rho_Q=0.4\,\rho_B .
 \label{eq:constraints}
\end{equation}
The auxiliary chemical potentials are fixed by these constraints within the model. Each temperature scan uses a solution family with fixed $\bm\mu$.

\begin{figure}[t]
\centering
\begin{minipage}[t]{0.49\textwidth}
\centering
\textbf{(a)}\par\smallskip
\includegraphics[width=\linewidth,height=0.28\textheight]{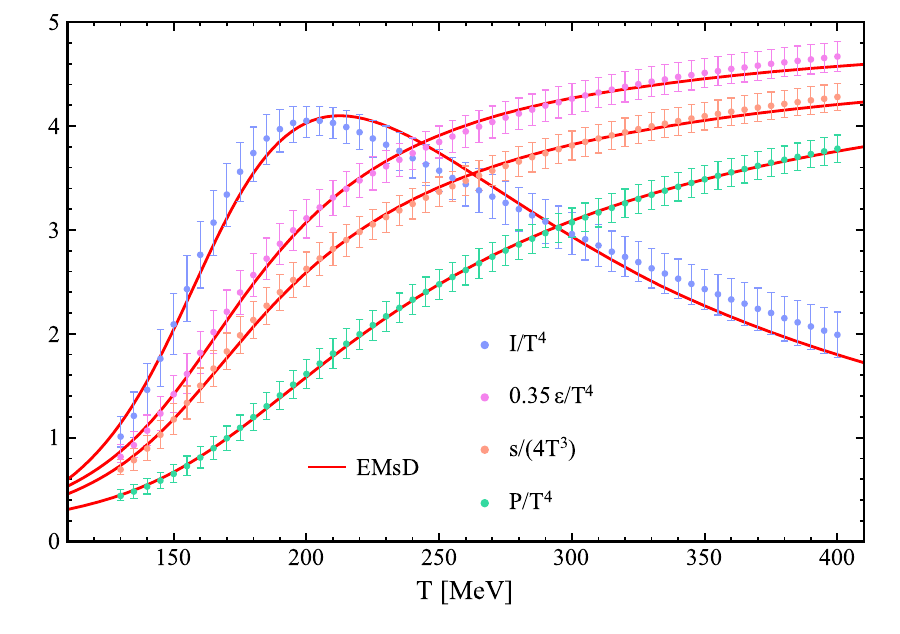}
\end{minipage}\hfill
\begin{minipage}[t]{0.49\textwidth}
\centering
\textbf{(b)}\par\smallskip
\includegraphics[width=\linewidth,height=0.28\textheight]{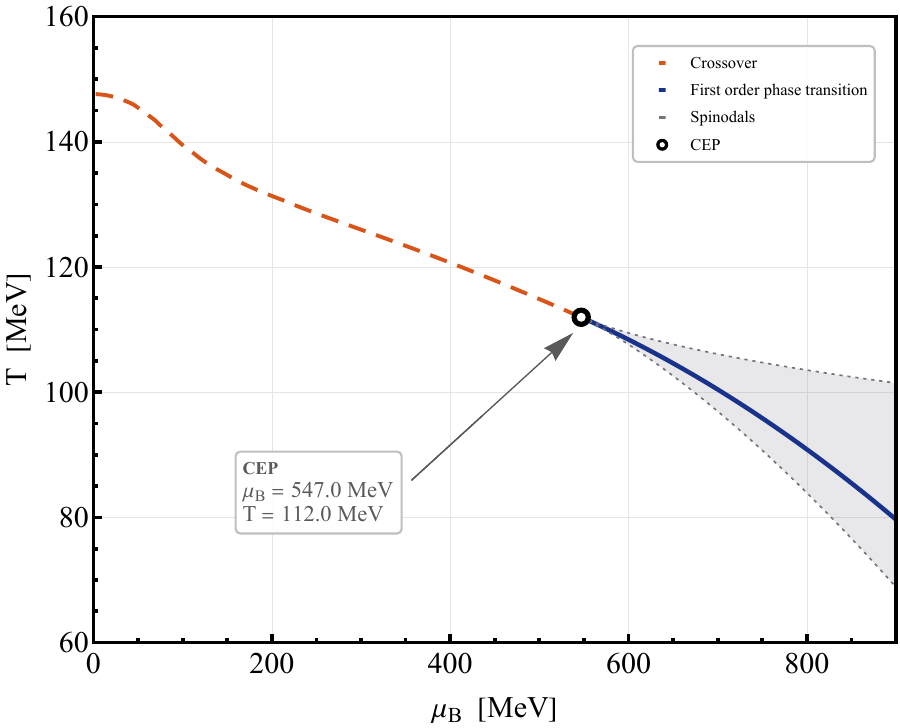}
\end{minipage}
\caption{Equation of state and phase diagram of the EMsD model. 
(a) Thermodynamics near zero density: red curves show $I/T^4=(\varepsilon-3p)/T^4$, $0.35\,\varepsilon/T^4$, $s/(4T^3)$, and $p/T^4$. Points with error bars are the $(2+1)$-flavor continuum lattice results of HotQCD~\cite{HotQCD:2014kol}; the factors $0.35$ and $1/4$ rescale both central values and errors for presentation.
(b) Phase diagram under $\rho_S=0$ and $\rho_Q/\rho_B=0.4$. The dashed orange and solid blue curves denote the crossover and first-order transition, respectively; the dotted gray curves bound the shaded spinodal region. The open circle marks the CEP at $(\mu_B,T)=(547.0,112.0)$~MeV.}
\label{fig:model}
\end{figure}

Fig.~\ref{fig:model}(a) compares the model equation of state with continuum lattice
results in the zero-density limit, $\mu_B\to0$. The pressure, entropy density, energy density,
and trace anomaly agree with the HotQCD results across the crossover and at higher
temperatures~\cite{HotQCD:2014kol}. The smooth temperature dependence of the pressure and
entropy density is consistent with the crossover nature of the transition at vanishing baryon
density. The tabulated observables are obtained from the same black-brane solutions with the
parameter set in Tab.~\ref{tab:model-parameters} and satisfy the thermodynamic identities
given above.

Finite-density tests of the same model were reported in
Ref.~\cite{Li:2026lxx}. Comparisons with lattice-QCD results along
fixed-$\mu_B/T$ trajectories showed agreement for the pressure and
charge densities at $\mu_Q=\mu_S=0$, and for the pressure, energy
density, and baryon density under $\mu_Q=0$ and $\rho_S=0$, up to
$\mu_B/T=3.5$. These comparisons used the zero-density parameter
set without additional finite-density fitting.

The phase diagram under the heavy-ion constraints is shown in Fig.~\ref{fig:model}(b).
Under these charge constraints, the model has a CEP at
\begin{equation}
 \mu_B^{\mathrm{CEP}}\simeq547.0\ \text{MeV},\qquad
 T^{\mathrm{CEP}}\simeq112.0\ \text{MeV}.
 \label{eq:cep}
\end{equation}
This point marks the termination of the first-order transition line. The low- and high-entropy
branches become indistinguishable there, and the entropy discontinuity between the coexisting
phases vanishes.

The gray shaded region between the spinodal curves contains a thermally unstable black-brane
branch, characterized by $(\partial T/\partial s)_{\bm\mu}<0$.
Multiple solution branches can occur at the same $(T,\bm\mu)$ within this region, and the
first-order coexistence line lies between the spinodal limits. The two spinodal curves meet at
the CEP.

\section{Temperature cumulants}
\label{sec:cumulants}

\subsection{Definition and evaluation of temperature cumulants}
\label{sec:numerics}

We consider equilibrium temperature fluctuations at fixed volume $V$ and fixed chemical potentials $\bm\mu=(\mu_B,\mu_Q,\mu_S)$. 
For a system with conserved charges $i\in\{B,Q,S\}$, the grand potential obeys
\begin{equation}
 d\Omega=-S\,dT-p\,dV-\sum_iN_i\,d\mu_i .
\end{equation}
A Legendre transformation with respect to the conjugate pair $(T,S)$ defines
\begin{equation}
 W=\Omega+TS=U-\sum_i\mu_iN_i,
 \qquad dW=T\,dS-p\,dV-\sum_iN_i\,d\mu_i .
\end{equation}
The corresponding densities $w=W/V=-p+Ts$ and $s=S/V$ satisfy $(\partial w/\partial s)_{\bm\mu}=T$. 
Following the thermodynamic framework of Ref.~\cite{Chen:2025vwl}, we define the dimensionless temperature cumulants by
\begin{equation}
 c_n=T^{3n-4}\left.\frac{\partial^n w}{\partial s^n}\right|_{\bm\mu}
     =T^{3n-4}\left.\frac{\partial^{n-1}T}{\partial s^{n-1}}\right|_{\bm\mu},
 \qquad n\ge2 .
 \label{eq:cn-definition}
\end{equation}
Thus the cumulant hierarchy is determined by successive entropy derivatives of the equation of state $T(s,\bm\mu)$.

For $\delta T=T-\langle T\rangle$, the physical cumulants are related to $c_n$ by
\begin{equation}
 \kappa_n(\delta T)=\frac{c_n\,T^n}{(VT^3)^{n-1}} .
 \label{eq:kappa}
\end{equation}
In particular, $\kappa_2=\langle(\delta T)^2\rangle$, while $\kappa_4$ is the connected fourth moment, $\langle(\delta T)^4\rangle-3\langle(\delta T)^2\rangle^2$.
Eq.~\eqref{eq:kappa} specifies the volume normalization used throughout this work.

For evaluation from the equation of state, it is also useful to introduce the dimensionless thermal susceptibilities
\begin{equation}
 \chi_n(T,\bm\mu)=T^{n-4}
 \left.\frac{\partial^n p}{\partial T^n}\right|_{\bm\mu},
 \label{eq:chiT}
\end{equation}
which, using $s=(\partial p/\partial T)_{\bm\mu}$, become
\begin{equation}
 \chi_n=T^{n-4}\left.\frac{\partial^{n-1}s}{\partial T^{n-1}}\right|_{\bm\mu} .
 \label{eq:chin-entropy}
\end{equation}
Inverse-function differentiation of Eq.~\eqref{eq:cn-definition} then gives
\begin{equation}
 c_2=\frac{T^2}{s'},\qquad
 c_3=-\frac{T^5s''}{s'^3},\qquad
 c_4=\frac{T^8\left(3s''^2-s's'''\right)}{s'^5},
 \label{eq:c234-entropy}
\end{equation}
where primes denote temperature derivatives at fixed $\bm\mu$. 
The fifth- and sixth-order cumulants are likewise expressed in terms of entropy-density derivatives as
\begin{align}
 c_5&=\frac{T^{11}}{(s')^7}
 \left[-15(s'')^3+10s's''s'''-(s')^2s^{(4)}\right],\nonumber\\
 c_6&=\frac{T^{14}}{(s')^9}
 \Bigl[105(s'')^4-105s'(s'')^2s'''+10(s')^2(s''')^2\nonumber\\
 &\hspace{5em}+15(s')^2s''s^{(4)}-(s')^3s^{(5)}\Bigr],
 \label{eq:c56-entropy}
\end{align}
with $s^{(k)}=(\partial^k s/\partial T^k)_{\bm\mu}$. 
Thus all cumulants through sixth order can be evaluated from the entropy density and its first five temperature derivatives~\cite{Chen:2025vwl}.

In the holographic model, $s$ is obtained directly from the horizon area~\cite{DeWolfe:2010he}. 
To cross-check the numerical evaluation, we calculate the cumulants using two representations of the same equation of state. 
In pipeline (i), we fit $s(T)$ locally and evaluate its temperature derivatives; in pipeline (ii), we fit $T(u)$ with $u=\ln(s/s_{\rm ref})$, where $s_{\rm ref}$ is an arbitrary fixed reference density. 
The operator identity
\begin{equation}
 \left.\frac{\partial}{\partial s}\right|_{\bm\mu}
 =\frac1s\left.\frac{\partial}{\partial u}\right|_{\bm\mu}
 \label{eq:entropy-operator}
\end{equation}
connects the latter parameterization directly to Eq.~\eqref{eq:cn-definition}. The coordinate $u$ remains regular as a branch parameter when $T(s)$ develops turning points, where a single fit to $s(T)$ cannot represent all branches. 
We compare $c_2$--$c_4$ obtained from the two representations wherever both evaluations are reliable, and vary the fitting windows and polynomial degrees to assess sensitivity to numerical differentiation. 
This comparison checks the consistency of the cumulants under a change of parameterization. 
Pipeline (i) is used for the temperature slices in Sec.~\ref{sec:slices}, while pipeline (ii) is used for the full plane and higher orders, for which the branch parameterization is better suited.
Both numerical routes evaluate derivatives within solution families with fixed $\bm\mu$.
Sampling the resulting cumulants along freeze-out is a subsequent operation, not differentiation along that trajectory.

These definitions apply to smooth single-phase branches. 
In the first-order region, stable and metastable branches can each have well-defined local derivatives, whereas an unstable branch cannot represent equilibrium fluctuations. 
At the coexistence line the two phases have equal pressure, and derivatives of the equilibrium equation of state require a separate treatment of its discontinuity. 
For the plane plots we conservatively mask the entire multibranch band between the spinodals ($3.7\%$ of the plotted plane). 
This plotting choice does not imply that all single-phase branch cumulants are undefined inside the band; we do not calculate mixed-phase cumulants here.

\subsection{Behavior at the CEP and in the conformal limit}
\label{sec:properties}

At the CEP, the enhanced thermal response suppresses temperature fluctuations. 
On a stable single-phase branch,
\begin{equation}
 \chi_2=\frac{s'}{T^2}>0,\qquad c_2=\frac{T^2}{s'},
 \label{eq:chi2pos}
\end{equation}
and the physical variance is
\begin{equation}
 \langle(\delta T)^2\rangle=\frac{c_2}{VT}=\frac{T^2}{C_{\bm\mu}},
 \qquad C_{\bm\mu}=VTs' .
 \label{eq:variance-heat-capacity}
\end{equation}
Using the mean-field description of the CEP in classical holographic models~\cite{DeWolfe:2010he,Chen:2018vty}, we expand the local equation of state at fixed $\bm\mu=\bm\mu_c$, where $T_c$, $\bm\mu_c$, and $s_c$ denote the values at the CEP of Eq.~\eqref{eq:cep}, as $T=T_c+a_3y^3+a_4y^4+a_5y^5+\cdots$, with $y=s-s_c$ and $a_3>0$.
Eq.~\eqref{eq:cn-definition} gives
\begin{align}
 c_2&\longrightarrow0, &
 c_3&\longrightarrow0, &
 c_4&\longrightarrow6a_3T_c^8,\nonumber\\
 c_5&\longrightarrow24a_4T_c^{11}, &
 c_6&\longrightarrow120a_5T_c^{14} .
 \label{eq:cep-cumulant-limits}
\end{align}
Thus $c_2$ and $c_3$ vanish at the CEP, with $c_3$ approaching zero from negative values below $T_c$ and positive values above it. 
The fourth-order cumulant tends to a finite positive constant. 
The fifth and sixth orders also remain finite, but their values and signs depend on the subleading terms of the equation of state; both vanish in the purely cubic approximation.

In the conformal limit, we take $s=A T^3$ with constant $A>0$. Eq.~\eqref{eq:cn-definition} gives
\begin{align}
 c_2&=\frac{1}{3A}, &
 c_3&=-\frac{2}{9A^2}, &
 c_4&=\frac{10}{27A^3},\nonumber\\
 c_5&=-\frac{80}{81A^4}, &
 c_6&=\frac{880}{243A^5} .
 \label{eq:conformal-signs}
\end{align}
Thus the even-order cumulants are positive and the odd-order ones negative. 
Following Ref.~\cite{Chen:2025vwl}, we define the normalized ratios
\begin{equation}
 R_{n2}\equiv\frac{c_n}{c_2^{n-1}},\qquad
 \frac{\kappa_n(\delta T)}{\kappa_2(\delta T)^{n-1}}
 =\frac{R_{n2}}{T^{n-2}}.
 \label{eq:Rn2-volume}
\end{equation}
The conformal reference values are
\begin{equation}
 R_{32}=-2,\qquad R_{42}=10,\qquad R_{52}=-80,\qquad R_{62}=880 .
 \label{eq:conformal-ratios}
\end{equation}
These provide a common reference for identifying departures from conformal behavior and sign reversals in the transition region.

\section{Results for \texorpdfstring{$c_2$, $c_3$, and $c_4$}{c2, c3 and c4}}
\label{sec:results}

We first examine the temperature dependence of $c_2$--$c_4$ at fixed chemical potentials and relate their characteristic structures to the thermal response of the medium. 
We then follow these structures over the $T$--$\muB$ plane and evaluate the cumulants along chemical freeze-out trajectories to determine their collision-energy dependence.

\subsection{Temperature dependence and thermal response}
\label{sec:slices}

\begin{figure}[H]
\centering
\includegraphics[width=\textwidth]{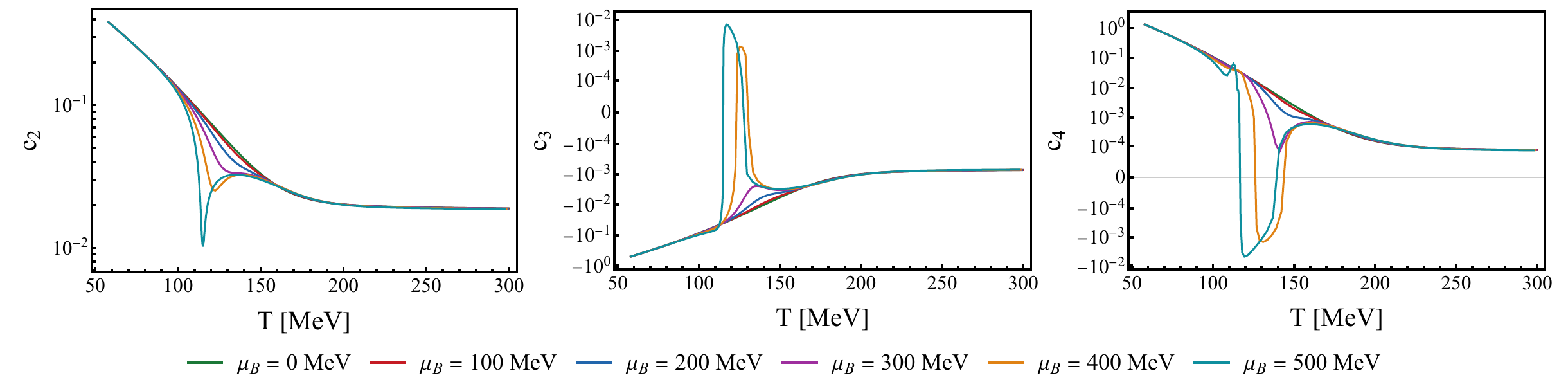}
\caption{Temperature cumulants $c_2$ (left), $c_3$ (middle), and $c_4$ (right) at six fixed baryon chemical potentials, labelled by the rounded values
$\muB=0$, $100$, $200$, $300$, $400$, and $500$~MeV. Each temperature scan keeps $(\mu_B,\mu_Q,\mu_S)$ fixed, with the auxiliary chemical potentials chosen to satisfy Eq.~\eqref{eq:constraints}.}
\label{fig:slices}
\end{figure}

Fig.~\ref{fig:slices} shows $c_2$, $c_3$, and $c_4$ as functions of temperature for $\muB=0$, $100$, $200$, $300$, $400$, and $500$~MeV. 
At low temperature, $T\lesssim75$~MeV, their dependence on $\muB$ is weak and the curves nearly coincide. 
As the temperature rises, $c_2$ decreases and approaches a plateau. 
At $T\simeq290$~MeV, $c_2$ is approximately one twentieth of its value at $T=58$~MeV, indicating weaker temperature fluctuations at high temperature. 
This behavior agrees qualitatively with the fRG results of Ref.~\cite{Chen:2025vwl}. 
At $T\simeq290$~MeV the three cumulants approach $c_2=1.89\times10^{-2}$, $c_3=-7.45\times10^{-4}$, and $c_4=7.37\times10^{-5}$, with only a few-percent variation across the chemical potentials shown. 
The common plateaus and the signs $c_3<0$, $c_4>0$ are consistent with the approach to the conformal regime discussed in Sec.~\ref{sec:properties}.

The dependence on $\muB$ is most pronounced in the transition region, $T\simeq110$--$150$~MeV. 
For the curves with $\muB\le300$~MeV, $c_3$ remains negative and $c_4$ positive throughout the temperature scan. At $\muB=400$ and $500$~MeV, intervals with $c_3>0$ and $c_4<0$ emerge. 
At $\muB=500$~MeV, the positive maximum of $c_3$ is about $7.1\times10^{-3}$ and the negative minimum of $c_4$ about $-4.5\times10^{-3}$; $c_2$ also develops a local minimum of approximately $1.02\times10^{-2}$. 
Increasing $\muB$ therefore introduces a localized suppression of $c_2$ and sign reversals of $c_3$ and $c_4$ on top of their smooth temperature dependence away from the transition.

The inverse-function relations in Eq.~\eqref{eq:c234-entropy} connect these features to the thermal response $s'(T)=C_{\bm\mu}/(VT)$~\cite{Chen:2025vwl,Liu:2025cpj}.
Since $c_2=T^2/s'$, an enhanced response suppresses $c_2$ at a given temperature.
The signs of the higher cumulants also depend on the shape of that response. 
On a stable branch, where $s'>0$,
\begin{align}
 \operatorname{sign}(c_3) & =-\operatorname{sign}(s''), \label{eq:signc3} \\
 c_4<0\quad & \Longleftrightarrow\quad s's'''>3(s'')^2.
 \label{eq:signc4}
\end{align}
Thus $c_3>0$ requires the thermal response to decrease with temperature, $s''<0$.
By contrast, $c_4$ can become negative while the response is still increasing, provided its curvature satisfies Eq.~\eqref{eq:signc4}. 
In our numerical results, this curvature condition is met at a lower $\muB$ than the condition $s''<0$, so the negative-$c_4$ region appears before the positive-$c_3$ region as $\muB$ increases.
The three cumulants thus probe complementary aspects of the same thermal response.
The next subsection follows these features over the full $T$--$\muB$ plane and examines their relation to the phase boundary
and the CEP.

\subsection{Suppression and sign structure in the \texorpdfstring{$T$--$\muB$}{T-muB} plane}
\label{sec:plane}

\begin{figure}[H]
\centering
\includegraphics[width=\textwidth]{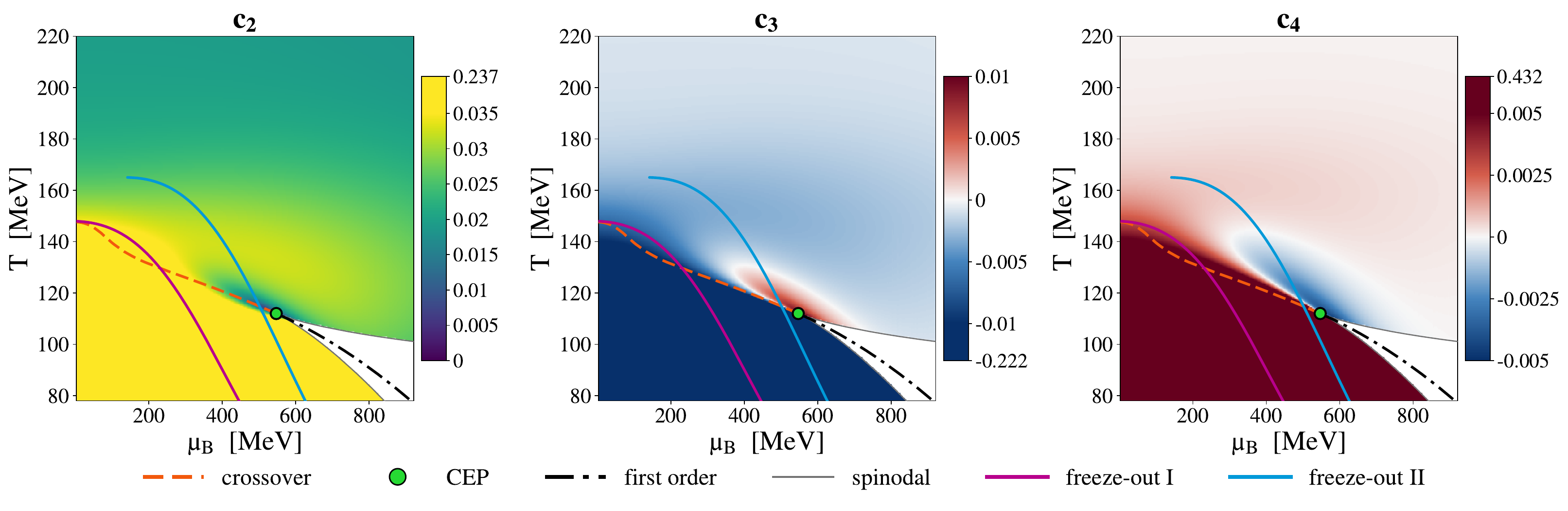}
\caption{Temperature cumulants $c_2$ (left), $c_3$ (middle), and $c_4$ (right) over the $T$--$\muB$ plane under the constraints $\rho_S=0$ and $\rho_Q/\rho_B=0.4$. The $c_2$ scale starts at zero, while the $c_3$ and $c_4$ scales are symmetric about zero. Curves mark the crossover (dashed), first-order transition (dash-dotted), and spinodals (thin solid); the circle marks the CEP. The two freeze-out trajectories correspond to $(T_{\rm lim},\mu_\Delta)=(148,-20)$~MeV (I) and $(165,130)$~MeV (II), with the parameterization given below. The blank wedge is the masked multibranch band between the spinodals; mixed-phase cumulants are not evaluated.}
\label{fig:plane}
\end{figure}

Fig.~\ref{fig:plane} displays $c_2$, $c_3$, and $c_4$ together with the phase boundary and the CEP. 
The multibranch band between the spinodals is masked, so the coloured regions represent smooth single-phase branches; mixed-phase fluctuations are not considered.

The left panel shows how the critical suppression of $c_2$ emerges from its smooth thermal background. 
At high temperature, $c_2$ is already small and varies weakly across the plane, consistent with the common plateau in the temperature scans. 
Near the CEP, a further localized depression develops close to the transition line. 
The inverse relation $c_2=1/\chi_2$ links this depression to the enhanced thermal susceptibility~\cite{Chen:2025vwl}. 
Its divergence at the CEP drives $c_2$ to zero, consistent with the variance suppression found in PNJL calculations~\cite{Liu:2026pro}.
The critical feature is therefore the additional suppression localized around the CEP relative to the surrounding background. 
A small value of $c_2$ by itself is also compatible with the smooth high-temperature regime.

The middle and right panels reveal extended regions with $c_3>0$ and $c_4<0$, both on the high-temperature side of the transition line. 
These regions begin at different chemical potentials: $c_4$ first becomes negative at $\muB\simeq305$~MeV, whereas $c_3$ first becomes positive at $\muB\simeq372$~MeV. 
Both sign reversals thus develop on the smooth crossover side, well before the critical chemical potential of $547$~MeV is reached. 
Their temperature ranges differ as well. 
Around $\muB\simeq500$~MeV, where the crossover occurs near $115$~MeV, the positive-$c_3$ interval extends approximately from $115$ to $127$~MeV and the negative-$c_4$ interval from $117$ to $139$~MeV. 
Positive $c_3$ appears closer to the transition line, while negative $c_4$ extends farther toward high temperature. 
The two regions overlap, but neither simply contains the other.

These distributions show that the sign changes characterize the evolving shape of the equation of state throughout part of the crossover region. A positive $c_3$ marks positive curvature of $T(s)$, while a negative $c_4$ marks a decrease of that curvature with entropy. 
The two conditions can hold over different intervals, allowing the cumulants to resolve different aspects of the same transition. 
Neither sign reversal requires a first-order transition, and neither sign boundary uniquely locates the CEP.
In the mean-field description appropriate to the classical holographic CEP~\cite{DeWolfe:2010he,Chen:2018vty}, Eq.~\eqref{eq:cep-cumulant-limits} gives a finite positive $c_4$ at the critical point itself. 
This local limit is distinct from the extended negative-$c_4$ region on its high-temperature side.

The separation between the $c_2$ depression and the two sign-changing regions has a direct consequence for freeze-out. 
A trajectory may sample reduced $c_2$ without entering either sign-changing region, or cross only one of them. 
The two trajectories shown in Fig.~\ref{fig:plane} illustrate the dependence on the path: freeze-out I avoids both sign-changing regions, whereas freeze-out II intersects both. 
Thus a single equation of state can yield different combinations of variance suppression and sign reversals along different trajectories. 
The collision-energy dependence is determined by how the freeze-out path samples these structures, which motivates the analysis below.

\subsection{Temperature cumulants along chemical freeze-out}
\label{sec:freezeout}

\begin{figure}[H]
\centering
\includegraphics[width=\textwidth]{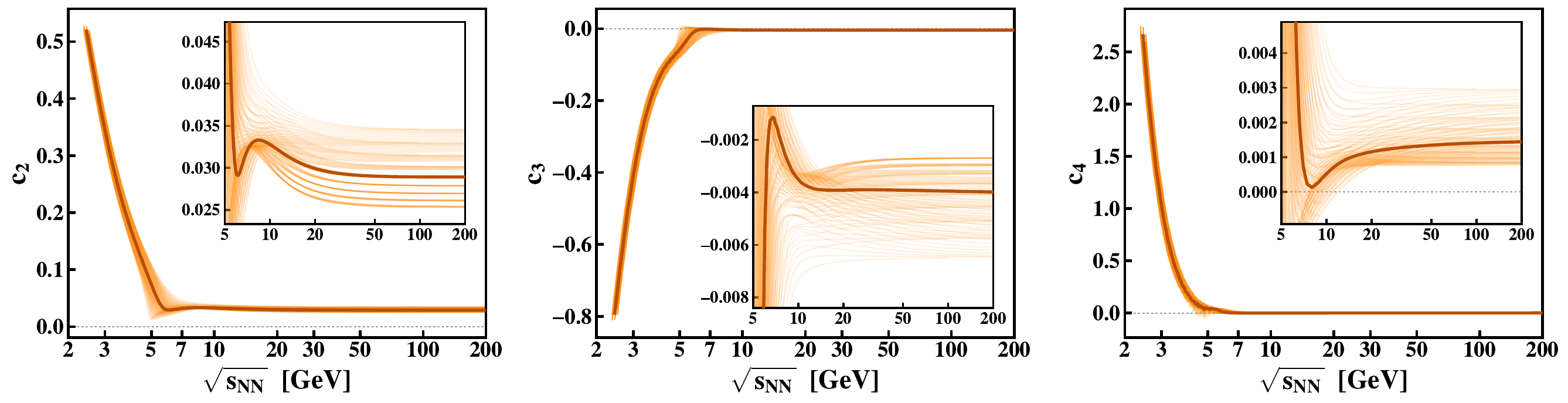}
\caption{Temperature cumulants $c_2$ (left), $c_3$ (middle), and $c_4$ (right) evaluated along chemical freeze-out trajectories as functions of collision energy $\snn$. The $117$ curves sample $T_{\rm lim}=148$--$165$~MeV and $\mu_\Delta=-20$--$130$~MeV on a $9\times13$ grid in Eq.~\eqref{eq:freezeout}; the dark curve corresponds to the midpoint $(T_{\rm lim},\mu_\Delta)=(156.5,55)$~MeV. The envelope represents a parameter scan, not a statistical confidence band. Insets show $\snn\ge5$~GeV.}
\label{fig:freezeout}
\end{figure}

To study how freeze-out conditions affect the imprint of the CEP, we compare $c_2$--$c_4$ along a family of trajectories. 
The equation of state and the CEP remain fixed throughout this comparison; only the path through the phase plane changes. 
The sensitivity of critical fluctuation signals to the location of freeze-out relative to the CEP motivates such comparisons~\cite{Stephanov:1999zu,Karthein:2025hvl}. 
We use the freeze-out parameterization of Ref.~\cite{Li:2026lxx}, based on hadron resonance gas fits and generalizing Refs.~\cite{Andronic:2017pug,Adamczyk:2017iwn},
\begin{align}
 \muB&=\frac{a}{1+b\,\snn}+\mu_\Delta ,\nonumber\\
 T&=\frac{T_{\rm lim}}{1+\exp\!\left[c-\ln\!\left(d\,\snn+e\right)/0.45\right]} ,
 \label{eq:freezeout}
\end{align}
where $a=1785$~MeV, $b=0.77$, $c=3.5$, $d=1.54$, and $e=-0.6$, with $\snn$ inserted in GeV.
We scan $T_{\rm lim}=148$--$165$~MeV and $\mu_\Delta=-20$--$130$~MeV on a $9\times13$ grid, with the two corner trajectories I and II shown in Fig.~\ref{fig:plane}. 
At each freeze-out point, the auxiliary chemical potentials are chosen to satisfy Eq.~\eqref{eq:constraints}.
The cumulants are calculated at fixed $\bm\mu$ before being evaluated along the trajectories. 
All sampled points lie outside the masked multibranch band.

Fig.~\ref{fig:freezeout} shows the resulting energy dependence. 
The cumulants have large magnitudes at low collision energies, where freeze-out samples the low-temperature equation of state. 
At high energies, the conditions approach $T=T_{\rm lim}$ and $\muB=\mu_\Delta$, and each cumulant tends to a plateau. 
Between these regimes, the insets reveal local extrema superimposed on the broader energy dependence.

The midpoint trajectory, $(T_{\rm lim},\mu_\Delta)=(156.5,55)$~MeV, serves as a reference. Along this line, $c_2$ reaches a local minimum of $0.0291$ at $\snn=6.1$~GeV, close to the intersection with the transition line at $5.9$~GeV. 
It then rises to a local maximum of approximately $0.0333$ at $\snn\simeq8.4$~GeV before approaching its high-energy plateau.
Relative to this maximum, the dip has a depth of about $13\%$.
The same trajectory gives a local maximum of $c_3$ at $6.8$~GeV and a local minimum of $c_4$ at $8.0$~GeV, although $c_3$ remains negative and $c_4$ positive.
We interpret this nonmonotonic pattern as an imprint of the CEP on the surrounding equation of state. 
Such an imprint can be sampled at a finite separation from the CEP without entering the sign-changing regions. 
The three extrema occur at different energies because the cumulants respond to different aspects of the thermodynamic structure encountered along the path.

Varying $T_{\rm lim}$ and $\mu_\Delta$ changes how the freeze-out line passes through these structures. 
Across the scan, the crossover-crossing energy ranges from about $5$ to $8$~GeV, while the local extrema move and change amplitude. Each trajectory also has its own high-energy plateau, set by its limiting freeze-out conditions. 
Local peak-to-dip contrasts, such as the relative depth quoted above for $c_2$, characterize these nonmonotonic structures more directly than the overall height of a curve.
The family of curves illustrates how the CEP imprint depends on the freeze-out path within the same model.

For some trajectories, the changes extend to sign reversals. Freeze-out I avoids both sign-changing regions in Fig.~\ref{fig:plane}, whereas freeze-out II crosses both.
In the full scan, $52$ of the $117$ trajectories enter at least one of these regions, including all with $\mu_\Delta\gtrsim105$~MeV. 
Thus local extrema can occur along paths that retain the background signs, while sign reversals identify paths that enter the corresponding lobes. 
The higher-order sign changes at low energy are treated qualitatively because they are more sensitive to numerical differentiation. 
This connection to the phase plane explains why the same equation of state produces different combinations of extrema and signs along different freeze-out lines.

The joint behavior of $c_2$, $c_3$, and $c_4$ provides complementary information about the relation between freeze-out and the structures around the CEP. 
The second order tracks the suppression of the variance, while the third and fourth orders resolve changes in the shape of the temperature fluctuations. 
A similar sensitivity of the $c_2$ dip to the relative position of freeze-out and the CEP has been reported in a PNJL calculation~\cite{Liu:2026pro}. 
The analysis below extends the study to higher orders and cumulant ratios, and examines the critical scaling of the hierarchy.

\section{Higher orders, cumulant ratios, and critical behavior}
\label{sec:higher}

We now extend the calculation to $c_5$ and $c_6$.
We first examine their temperature dependence, sign structure, and freeze-out projection.
We then study the normalized ratios $R_{n2}$ and use cross-order combinations to estimate the critical exponent $\delta$.

\subsection{Fifth- and sixth-order cumulants}
\label{sec:c56}
\label{sec:c56plane}
\label{sec:c56freezeout}

\begin{figure}[!htbp]
\centering
\includegraphics[width=\textwidth]{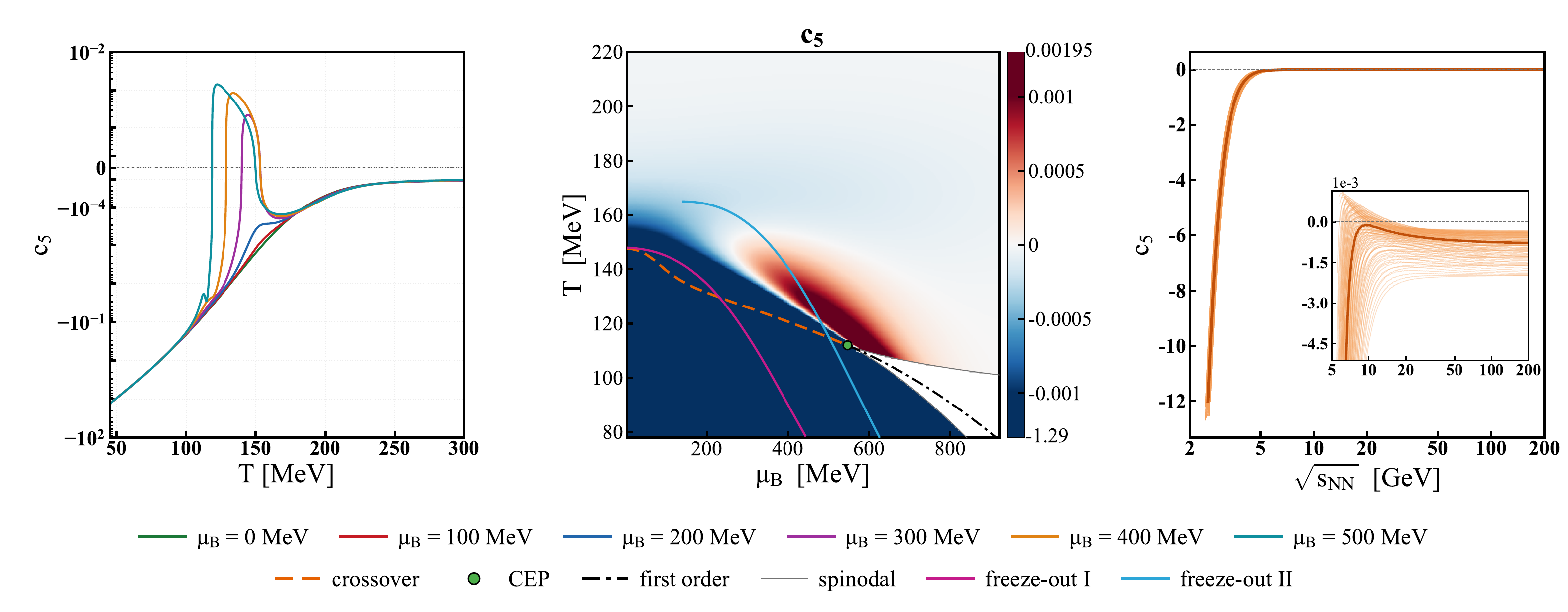}
\caption{The fifth-order temperature cumulant $c_5$.
Left: temperature dependence at the six fixed chemical potentials used in Fig.~\ref{fig:slices}; the vertical axis is symmetric-logarithmic.
Middle: the $T$--$\muB$ plane, with the phase boundaries, CEP, and freeze-out trajectories of Fig.~\ref{fig:plane}.
Right: the freeze-out family of Eq.~\eqref{eq:freezeout}, sampled on the same $9\times13$ grid in $(T_{\rm lim},\mu_\Delta)$ as Fig.~\ref{fig:freezeout}, giving $117$ trajectories at $300$ collision energies.
The dark curve is the midpoint trajectory. The family represents a parameter scan, not a confidence band.
The inset shows $\snn\ge5$~GeV.}
\label{fig:c5}
\end{figure}

\begin{figure}[!htbp]
\centering
\includegraphics[width=\textwidth]{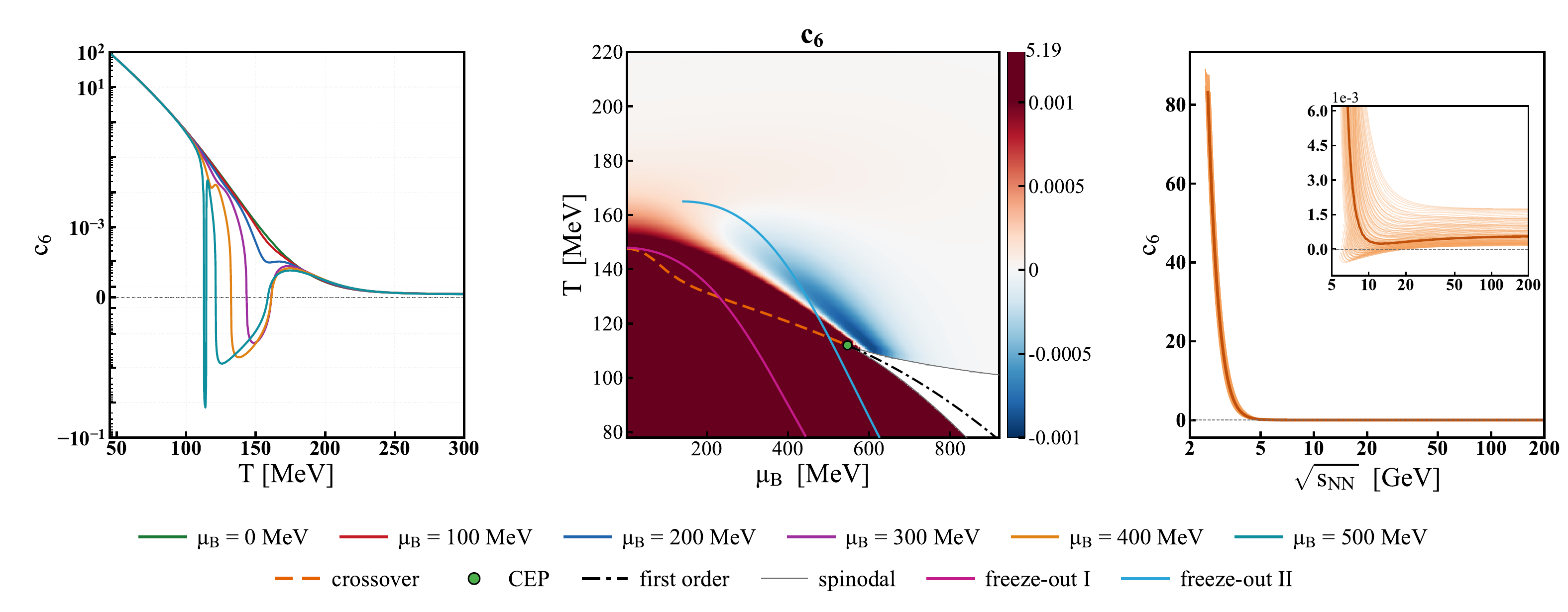}
\caption{The sixth-order temperature cumulant $c_6$, with the same panel arrangement, chemical potentials, phase boundaries, and freeze-out family as Fig.~\ref{fig:c5}.}
\label{fig:c6}
\end{figure}

Figs.~\ref{fig:c5} and~\ref{fig:c6} display $c_5$ and $c_6$ in three complementary ways.
The left panels show their temperature dependence at fixed $\bm\mu$.
At low temperature, $c_5$ is negative and $c_6$ positive, with large magnitudes and weak dependence on $\muB$.
Their magnitudes decrease as the temperature rises and approach high-temperature plateaus.
Sign reversals appear in the transition region: the $\muB=300$~MeV curves already show $c_5>0$ and $c_6<0$.
These reversals therefore occur at lower chemical potentials than those of $c_3$ and $c_4$.

The middle panels show the regions with $c_5>0$ and $c_6<0$ in the $T$--$\muB$ plane.
These regions lie on the high-temperature side of the phase boundary and correspond to the sign reversals in the left panels.
The positive-$c_5$ region begins at $\muB\simeq264$~MeV, and the negative-$c_6$ region at $\muB\simeq238$~MeV.
Together with Sec.~\ref{sec:plane}, the onset chemical potentials obey
\begin{equation}
 \muB^{(6)}<\muB^{(5)}<\muB^{(4)}<\muB^{(3)}:
 \qquad 238<264<305<372~\text{MeV}.
 \label{eq:onsets}
\end{equation}
Higher orders thus respond to changes in the shape of the thermal response at lower $\muB$ in this model.
Their early onsets reflect the sensitivity of higher-order cumulants to the crossover equation of state.

At the CEP itself, Eq.~\eqref{eq:cep-cumulant-limits} predicts finite $c_5$ and $c_6$ for an analytic mean-field equation of state.
Their limiting values and signs depend on its fourth- and fifth-order terms in $s-s_c$.
Here $c_5$ probes the coefficient $a_4$, while $c_6$ probes $a_5$ in the local expansion of $T(s)$.
Thus $c_5$ and $c_6$ can approach finite values at the CEP while changing sign at nearby temperatures.

The right panels evaluate the fixed-$\bm\mu$ cumulants along the freeze-out family introduced in Sec.~\ref{sec:freezeout}.
The large low-energy values reflect the low-temperature background, while each trajectory approaches its own high-energy plateau.
The insets use a narrower vertical scale to show the local extrema at $\snn\ge5$~GeV.
On the midpoint trajectory, $|c_5|$ and $|c_6|$ have minima near $9.5$ and $12.7$~GeV, respectively.
These positions differ from the lower-order extrema because each cumulant probes a different derivative of the same equation of state~\cite{Chen:2025vwl}.
Their collision-energy dependence also depends on the freeze-out path.
Changing the trajectory changes both the temperatures and chemical potentials sampled by the cumulants.
The spread of curves therefore shows how the same phase-plane structure is projected onto collision energy.

\subsection{Cumulant ratios}
\label{sec:ratios}

\begin{figure}[H]
\centering
\includegraphics[width=0.8\textwidth]{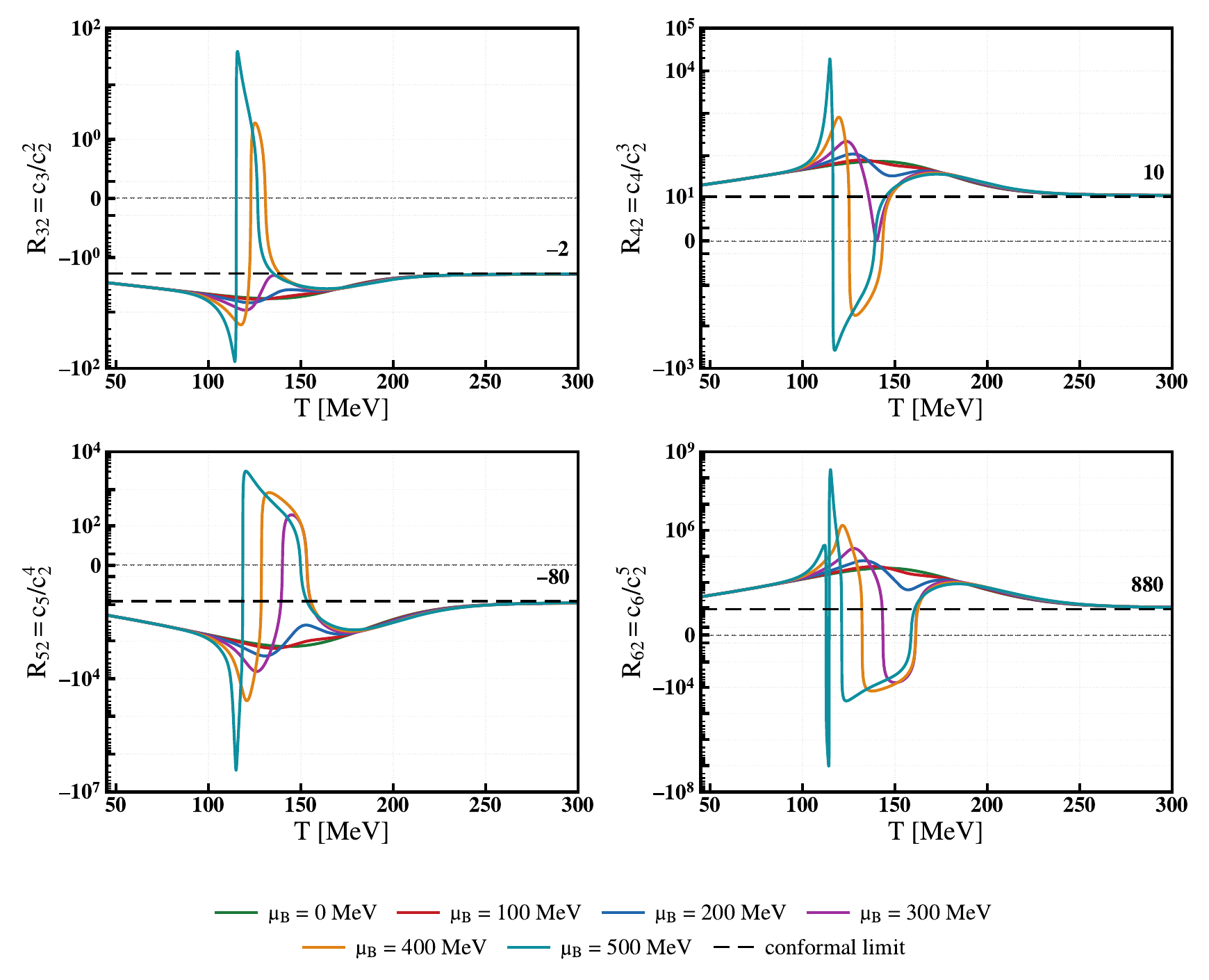}
\caption{Normalized cumulant ratios $R_{n2}=c_n/c_2^{\,n-1}$ for $n=3$, $4$, $5$, and $6$ at the six fixed chemical potentials of Fig.~\ref{fig:slices}.
The vertical axes are symmetric-logarithmic.
Black dashed lines mark the conformal limits $-2$, $10$, $-80$, and $880$; thin grey dashed lines mark zero.}
\label{fig:ratios}
\end{figure}

Fig.~\ref{fig:ratios} shows the volume-independent ratios defined in Eq.~\eqref{eq:Rn2-volume}~\cite{Chen:2025vwl}.
At high temperature they approach the conformal reference values of Eq.~\eqref{eq:conformal-ratios}.
At $T=299$~MeV and the smallest tabulated $\muB$, we obtain
\begin{equation}
 R_{32}=-2.076,\qquad R_{42}=10.82,\qquad R_{52}=-90.5,\qquad R_{62}=1047.
\end{equation}
The deviations from the conformal limits range from $3.8\%$ to $19.0\%$ and increase with order.
At this finite temperature, residual nonconformality is still visible in the ratios.

Near the transition, the ratios develop pronounced peaks and dips.
The suppression of $c_2$ amplifies their magnitudes through the denominator, while the numerator determines their signs.
For example, $R_{42}$ reaches about $1.7\times10^4$ at $\muB\simeq500$~MeV, compared with its conformal value of $10$.
A large ratio therefore need not imply a large numerator cumulant.
The ratios depend on how rapidly the thermal response changes with temperature relative to its own magnitude.
Contributions from its higher derivatives can reinforce or cancel one another, strengthening or weakening the peaks and dips.

The structures strengthen with $\muB$ in the slices shown here.
In comparison, the fRG and PNJL calculations report weaker broad peaks and dips at larger $\muB$~\cite{Chen:2025vwl,Liu:2025cpj}.
The PNJL study attributes these broad features to deconfinement and distinguishes them from oscillations near the chiral CEP~\cite{Liu:2025cpj}.
In our results, the growing peaks and dips occur near the transition as the slices approach the CEP from the crossover side.
This enhancement is consistent with the mean-field critical behavior of the classical-gravity model~\cite{DeWolfe:2010he,Chen:2018vty}.
At the CEP, $c_2$ vanishes while $c_4$ remains finite and positive, strongly enhancing $R_{42}$ through its denominator.
The peak heights at finite distance also depend on the thermal background and higher derivatives of the equation of state.

We next use cross-order cumulant ratios to extract the critical exponent $\delta$.
In the leading scaling form, these combinations cancel the equation-of-state amplitude and the dependence on distance to the CEP, leaving limits determined by $\delta$.

\subsection{Critical scaling and the exponent \texorpdfstring{$\delta$}{delta}}
\label{sec:scaling}
\label{sec:cn-scaling}
\label{sec:exponents}
\label{sec:invariants}
\label{sec:un}
\label{sec:delta}

We approach the CEP at fixed $\bm\mu=\bm\mu_c$.
We assume that both this path and the entropy have nonzero projections onto the ordering sector~\cite{Nonaka:2004pg,Hatta:2002sj}.
The leading scaling is then governed by the critical-isotherm exponent $\delta$.
With $x=(T-T_c)/T_c$, Eq.~\eqref{eq:cn-definition} gives
\begin{equation}
 |c_n^{\rm lead}|\propto |x|^{p_n},
 \qquad p_n=1-\frac{n-1}{\delta}.
 \label{eq:cn-scaling}
\end{equation}
Here $\chi_2\propto|x|^{-\epsilon}$ and $c_2\propto|x|^\epsilon$, with $\epsilon=1-1/\delta$ the path exponent rather than the thermal-axis exponent $\alpha$~\cite{Hatta:2002sj,DeWolfe:2010he}.
Eq.~\eqref{eq:cn-scaling} applies when the leading derivative coefficient is nonzero.

For the mean-field value $\delta=3$~\cite{DeWolfe:2010he,Chen:2018vty}, $c_2\propto|x|^{2/3}$, $c_3\propto\operatorname{sgn}(x)|x|^{1/3}$, and $c_4$ remains finite and positive.
The leading cubic term gives no $c_5$ or $c_6$ contribution.
Their finite limits follow from higher-order terms in Eq.~\eqref{eq:cep-cumulant-limits}.
For nonzero limiting $c_5$ and $c_6$, $|R_{52}|\propto|x|^{-8/3}$ and $|R_{62}|\propto|x|^{-10/3}$ because $c_2$ vanishes.
For the three-dimensional Ising value $\delta=4.789$~\cite{Pelissetto:2000ek}, Eq.~\eqref{eq:cn-scaling} gives $p_5=0.165$ and $p_6=-0.044$.
Thus the leading $c_5$ vanishes, while $c_6$ diverges weakly.

Cross-order ratios provide a direct test of the common exponent.
For the leading scaling terms, the combinations
\begin{equation}
 u_3\equiv\frac{c_4c_2}{c_3^2}
       \longrightarrow\frac{\delta-2}{\delta-1},
 \qquad
 u_4\equiv\frac{c_5c_3}{c_4^2}
       \longrightarrow\frac{\delta-3}{\delta-2}
 \label{eq:u3u4}
\end{equation}
remove both the equation-of-state amplitude and the critical distance.
At mean field, $c_3\to0$ while $c_4$ stays nonzero.
Thus $u_4\to0$ also holds for a finite nonzero $c_5$.
The targets are $(u_3,u_4)=(1/2,0)$ for mean field and $(0.736,0.641)$ for the Ising scaling form.
In $u_3$, the numerator and denominator vanish with the same leading power.
The ratio $u_4$ checks the same scaling with the next cumulant order.

\begin{figure}[!htbp]
\centering
\includegraphics[width=\textwidth]{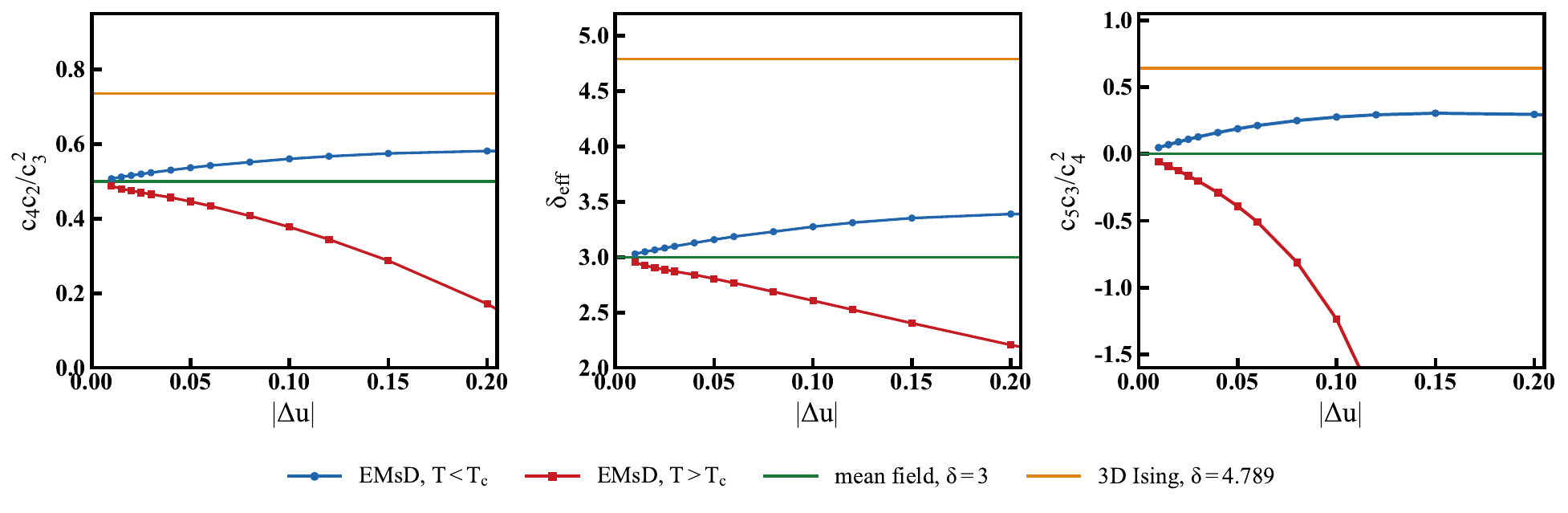}
\caption{$u_3$ (left), $\delta_{\rm eff}$ (middle), and $u_4$ (right) along the fixed-$\bm\mu_c$ path, versus $|\Delta u|$, with $\Delta u=u-u_c$ and $u=\ln(s/s_{\rm ref})$. Colors and reference limits are given in the legend.}
\label{fig:delta}
\end{figure}

We evaluate $u_3$ directly from $c_2$, $c_3$, and $c_4$ on both sides of the CEP.
Eq.~\eqref{eq:u3u4} gives
\begin{equation}
 \delta_{\rm eff}=\frac{2-u_3}{1-u_3},
 \qquad
 \lim_{T\to T_c^\pm}\delta_{\rm eff}=\delta.
 \label{eq:delta-from-u3}
\end{equation}
The two limits are taken along the same fixed-$\bm\mu_c$ path.
At finite distance, $\delta_{\rm eff}$ can differ between the two sides because of subleading terms in the equation of state, such as the $a_4$ term in the mean-field expansion above~\cite{Wegner:1972}.
These differences do not imply different critical exponents.

Fig.~\ref{fig:delta} shows how the two branches approach the CEP.
The ratio $u_3$ approaches $1/2$ from above for $T<T_c$ and from below for $T>T_c$.
The corresponding $\delta_{\rm eff}$ curves approach $3$.
The ratio $u_4$ approaches zero from opposite signs.
For example, at $|\Delta u|=0.01$ the two branches give $u_3\simeq0.507$ and $0.488$, and $u_4\simeq+0.047$ and $-0.059$, respectively.
Their common critical limits are consistent with
\begin{equation}
 \delta\simeq3.0,
 \label{eq:delta}
\end{equation}
as expected for the mean-field CEP of this model.
The ratios therefore connect the critical behavior of different cumulant orders through the same equation of state.

\section{Discussion and conclusions}
\label{sec:conclusions}

We have studied temperature cumulants through sixth order in the three-flavor multi-charge EMsD model under the heavy-ion constraints $\rho_S=0$ and $\rho_Q/\rho_B=0.4$~\cite{Li:2026lxx}.
The model has a CEP at $(\mu_B,T)\simeq(547,112)$~MeV, and the cumulants reveal how its thermal response evolves across the surrounding phase diagram.
The relation $c_2=T^2/s'$ connects the enhanced thermal response to a suppression of temperature fluctuations~\cite{Chen:2025vwl}.
Near the CEP, this suppression forms a localized depression against the smooth thermal background, while the higher cumulants resolve changes in the shape of the response.
Their reversed-sign regions develop on the high-temperature side of the transition and already extend into the smooth crossover region.
The onset chemical potentials are approximately $238$, $264$, $305$, and $372$~MeV for $c_6$, $c_5$, $c_4$, and $c_3$, respectively.
This ordering shows that higher orders detect changes in the crossover equation of state at lower chemical potentials in this model.
Positive $c_3$ marks a declining thermal response, whereas negative $c_4$ reflects sufficiently strong curvature of that response.
Together, the variance and higher cumulants describe both the strength and the shape of the thermal response.
The overall suppression and negative skewness away from the transition agree qualitatively with the fRG results of Ref.~\cite{Chen:2025vwl}, while the variance depression agrees with the PNJL study of Ref.~\cite{Liu:2026pro}.
Our results extend this picture by connecting the behavior through sixth order across a phase diagram subject to coupled conserved-charge constraints.

The critical analysis connects these different orders through a common local scaling form.
Along the fixed-$\bm\mu_c$ path, the critical-isotherm exponent $\delta$ controls the leading behavior when both the path and entropy have nonzero projections onto the ordering sector~\cite{Nonaka:2004pg,Hatta:2002sj}.
For the analytic mean-field equation of state, $c_2$ and $c_3$ vanish at the CEP, with $c_3$ approaching zero from opposite signs on its two sides.
The fourth order tends to a finite positive value, and the finite limits of $c_5$ and $c_6$ depend on higher-order terms beyond the leading cubic equation of state.
The onset of sign reversals in the crossover region and the limiting behavior at the CEP therefore probe different properties of the same equation of state.
Higher orders add sensitivity to its shape, while the normalized ratios $R_{n2}=c_n/c_2^{n-1}$ amplify the transition structure through the suppression of $c_2$.
The cross-order combinations $u_3$ and $u_4$ provide a further test by removing the leading amplitude and critical distance.
On both sides of the CEP they approach $(u_3,u_4)=(1/2,0)$, corresponding to $\delta\simeq3.0$ and consistent with mean-field criticality in classical holographic models~\cite{DeWolfe:2010he,Chen:2018vty}.
The differences between the two branches at finite distance reflect subleading terms in the equation of state and decrease toward the CEP.
These relations tie the variance suppression and higher-order limits to a common critical description.
The three-dimensional Ising exponents give different ratio benchmarks~\cite{Pelissetto:2000ek}, making the cross-order relations useful targets for future studies beyond the present approximation.

The freeze-out results show how the structures in the phase plane translate into collision-energy dependence.
On the midpoint trajectory, $c_2$ reaches a local minimum at $\snn=6.1$~GeV, close to the trajectory's crossing of the transition line.
The dip is about $13\%$ below the neighbouring local maximum and is accompanied by a $c_3$ maximum at $6.8$~GeV and a $c_4$ minimum at $8.0$~GeV.
Along this trajectory, the higher cumulants retain their background signs, whereas other trajectories enter the reversed-sign regions.
These different combinations of extrema and signs reflect how each freeze-out path samples the same equation of state.
The dip position and prominence thus carry information about the relative geometry of freeze-out and the thermal response around the CEP.
The quoted energies refer to the chosen model and reference trajectory; their shifts across the scan characterize the sensitivity to freeze-out conditions.
This path dependence also enters conserved-charge fluctuation studies~\cite{Karthein:2025hvl}, and a joint analysis of the temperature and charge channels could provide complementary constraints~\cite{Li:2026lxx}.
Temperature cumulants describe the thermal response, while charge cumulants probe derivatives with respect to chemical potentials.
Their combined energy dependence could help distinguish which parts of the phase diagram are sampled by a given freeze-out trajectory.
Mean-transverse-momentum fluctuations offer a possible experimental connection to the temperature channel~\cite{Gavin:2003cb,Gardim:2019xjs,Giacalone:2020lbm}.
Their measured cumulants also depend on geometry, collective flow, and volume fluctuations~\cite{Giacalone:2020lbm,ATLAS:2024jvf}, with further contributions from particle composition and hadronic dynamics~\cite{Reichert:2026ztj,Liu:2026lnr}.
The present equilibrium results identify thermodynamic structures whose survival and magnitude in these observables can be tested with dynamical calculations.

The following limitations set the scope of this interpretation.
Calibration to zero-density lattice data constrains the thermal background, while the finite-density phase structure remains model dependent.
The classical-gravity calculation describes a mean-field CEP and provides a benchmark for testing the cumulant relations in this limit.
The present numerical analysis supports this benchmark; distinguishing it reliably from non-mean-field criticality requires further validation of the ratio method.
The numerical resolution near the transition also limits the detailed structure of $c_5$ and $c_6$, while mixed-phase cumulants lie outside the smooth single-phase calculation.
Their analytic critical limits are therefore established separately from the features resolved in the numerical maps.
The freeze-out envelope represents a scan of trajectory parameters, so it illustrates path sensitivity rather than a statistical confidence region.
Finally, critical slowing down and the finite lifetime of the medium can modify equilibrium fluctuation patterns~\cite{Berdnikov:1999slowing,Mukherjee:2015swa,Mukherjee:2016kyu}.
A quantitative experimental comparison requires finite-volume dynamics, a mapping to transverse-momentum fluctuations, and the same acceptance as the measurement.
The most direct next steps are denser sampling near the CEP and tests of the cross-order ratios with noninteger critical exponents.
A comparison with $\mu_Q=\mu_S=0$ would also isolate the influence of the heavy-ion charge constraints.
Together with dynamical evolution, these studies would build on the equilibrium connection established here between thermal-response enhancement, variance suppression, and higher-order sign structure.

\begin{acknowledgments}
This work was supported in part by the Natural Science Foundation of Henan Province of China under Grant No. 262300421877, the National Natural Science Foundation of China (NSFC) Grant No. 12305136 and the start-up funding of Hangzhou Normal University under Grant No. 4245C50223204075.
\end{acknowledgments}

\bibliography{refs}

\begin{thebibliography}{67}%
\makeatletter
\providecommand \@ifxundefined [1]{%
 \@ifx{#1\undefined}
}%
\providecommand \@ifnum [1]{%
 \ifnum #1\expandafter \@firstoftwo
 \else \expandafter \@secondoftwo
 \fi
}%
\providecommand \@ifx [1]{%
 \ifx #1\expandafter \@firstoftwo
 \else \expandafter \@secondoftwo
 \fi
}%
\providecommand \natexlab [1]{#1}%
\providecommand \enquote  [1]{``#1''}%
\providecommand \bibnamefont  [1]{#1}%
\providecommand \bibfnamefont [1]{#1}%
\providecommand \citenamefont [1]{#1}%
\providecommand \href@noop [0]{\@secondoftwo}%
\providecommand \href [0]{\begingroup \@sanitize@url \@href}%
\providecommand \@href[1]{\@@startlink{#1}\@@href}%
\providecommand \@@href[1]{\endgroup#1\@@endlink}%
\providecommand \@sanitize@url [0]{\catcode `\\12\catcode `\$12\catcode
  `\&12\catcode `\#12\catcode `\^12\catcode `\_12\catcode `\%12\relax}%
\providecommand \@@startlink[1]{}%
\providecommand \@@endlink[0]{}%
\providecommand \url  [0]{\begingroup\@sanitize@url \@url }%
\providecommand \@url [1]{\endgroup\@href {#1}{\urlprefix }}%
\providecommand \urlprefix  [0]{URL }%
\providecommand \Eprint [0]{\href }%
\providecommand \doibase [0]{https://doi.org/}%
\providecommand \selectlanguage [0]{\@gobble}%
\providecommand \bibinfo  [0]{\@secondoftwo}%
\providecommand \bibfield  [0]{\@secondoftwo}%
\providecommand \translation [1]{[#1]}%
\providecommand \BibitemOpen [0]{}%
\providecommand \bibitemStop [0]{}%
\providecommand \bibitemNoStop [0]{.\EOS\space}%
\providecommand \EOS [0]{\spacefactor3000\relax}%
\providecommand \BibitemShut  [1]{\csname bibitem#1\endcsname}%
\let\auto@bib@innerbib\@empty
\bibitem [{\citenamefont {Aoki}\ \emph {et~al.}(2006)\citenamefont {Aoki},
  \citenamefont {Endrodi}, \citenamefont {Fodor}, \citenamefont {Katz},\ and\
  \citenamefont {Szabo}}]{Aoki:2006we}%
  \BibitemOpen
  \bibfield  {author} {\bibinfo {author} {\bibfnamefont {Y.}~\bibnamefont
  {Aoki}}, \bibinfo {author} {\bibfnamefont {G.}~\bibnamefont {Endrodi}},
  \bibinfo {author} {\bibfnamefont {Z.}~\bibnamefont {Fodor}}, \bibinfo
  {author} {\bibfnamefont {S.~D.}\ \bibnamefont {Katz}},\ and\ \bibinfo
  {author} {\bibfnamefont {K.~K.}\ \bibnamefont {Szabo}},\ }\bibfield  {title}
  {\bibinfo {title} {{The Order of the quantum chromodynamics transition
  predicted by the standard model of particle physics}},\ }\href
  {https://doi.org/10.1038/nature05120} {\bibfield  {journal} {\bibinfo
  {journal} {Nature}\ }\textbf {\bibinfo {volume} {443}},\ \bibinfo {pages}
  {675} (\bibinfo {year} {2006})},\ \Eprint
  {https://arxiv.org/abs/hep-lat/0611014} {arXiv:hep-lat/0611014} \BibitemShut
  {NoStop}%
\bibitem [{\citenamefont {Borsanyi}\ \emph {et~al.}(2014)\citenamefont
  {Borsanyi}, \citenamefont {Fodor}, \citenamefont {Hoelbling}, \citenamefont
  {Katz}, \citenamefont {Krieg},\ and\ \citenamefont
  {Szabo}}]{Borsanyi:2013bia}%
  \BibitemOpen
  \bibfield  {author} {\bibinfo {author} {\bibfnamefont {S.}~\bibnamefont
  {Borsanyi}}, \bibinfo {author} {\bibfnamefont {Z.}~\bibnamefont {Fodor}},
  \bibinfo {author} {\bibfnamefont {C.}~\bibnamefont {Hoelbling}}, \bibinfo
  {author} {\bibfnamefont {S.~D.}\ \bibnamefont {Katz}}, \bibinfo {author}
  {\bibfnamefont {S.}~\bibnamefont {Krieg}},\ and\ \bibinfo {author}
  {\bibfnamefont {K.~K.}\ \bibnamefont {Szabo}},\ }\bibfield  {title} {\bibinfo
  {title} {{Full result for the QCD equation of state with 2+1 flavors}},\
  }\href {https://doi.org/10.1016/j.physletb.2014.01.007} {\bibfield  {journal}
  {\bibinfo  {journal} {Phys. Lett. B}\ }\textbf {\bibinfo {volume} {730}},\
  \bibinfo {pages} {99} (\bibinfo {year} {2014})},\ \Eprint
  {https://arxiv.org/abs/1309.5258} {arXiv:1309.5258 [hep-lat]} \BibitemShut
  {NoStop}%
\bibitem [{\citenamefont {Bazavov}\ \emph {et~al.}(2014)\citenamefont {Bazavov}
  \emph {et~al.}}]{HotQCD:2014kol}%
  \BibitemOpen
  \bibfield  {author} {\bibinfo {author} {\bibfnamefont {A.}~\bibnamefont
  {Bazavov}} \emph {et~al.} (\bibinfo {collaboration} {HotQCD}),\ }\bibfield
  {title} {\bibinfo {title} {{Equation of state in ( 2+1 )-flavor QCD}},\
  }\href {https://doi.org/10.1103/PhysRevD.90.094503} {\bibfield  {journal}
  {\bibinfo  {journal} {Phys. Rev. D}\ }\textbf {\bibinfo {volume} {90}},\
  \bibinfo {pages} {094503} (\bibinfo {year} {2014})},\ \Eprint
  {https://arxiv.org/abs/1407.6387} {arXiv:1407.6387 [hep-lat]} \BibitemShut
  {NoStop}%
\bibitem [{\citenamefont {Stephanov}\ \emph {et~al.}(1998)\citenamefont
  {Stephanov}, \citenamefont {Rajagopal},\ and\ \citenamefont
  {Shuryak}}]{Stephanov:1998dy}%
  \BibitemOpen
  \bibfield  {author} {\bibinfo {author} {\bibfnamefont {M.~A.}\ \bibnamefont
  {Stephanov}}, \bibinfo {author} {\bibfnamefont {K.}~\bibnamefont
  {Rajagopal}},\ and\ \bibinfo {author} {\bibfnamefont {E.~V.}\ \bibnamefont
  {Shuryak}},\ }\bibfield  {title} {\bibinfo {title} {{Signatures of the
  tricritical point in QCD}},\ }\href
  {https://doi.org/10.1103/PhysRevLett.81.4816} {\bibfield  {journal} {\bibinfo
   {journal} {Phys. Rev. Lett.}\ }\textbf {\bibinfo {volume} {81}},\ \bibinfo
  {pages} {4816} (\bibinfo {year} {1998})},\ \Eprint
  {https://arxiv.org/abs/hep-ph/9806219} {arXiv:hep-ph/9806219} \BibitemShut
  {NoStop}%
\bibitem [{\citenamefont {Stephanov}\ \emph {et~al.}(1999)\citenamefont
  {Stephanov}, \citenamefont {Rajagopal},\ and\ \citenamefont
  {Shuryak}}]{Stephanov:1999zu}%
  \BibitemOpen
  \bibfield  {author} {\bibinfo {author} {\bibfnamefont {M.~A.}\ \bibnamefont
  {Stephanov}}, \bibinfo {author} {\bibfnamefont {K.}~\bibnamefont
  {Rajagopal}},\ and\ \bibinfo {author} {\bibfnamefont {E.~V.}\ \bibnamefont
  {Shuryak}},\ }\bibfield  {title} {\bibinfo {title} {{Event-by-event
  fluctuations in heavy ion collisions and the QCD critical point}},\ }\href
  {https://doi.org/10.1103/PhysRevD.60.114028} {\bibfield  {journal} {\bibinfo
  {journal} {Phys. Rev. D}\ }\textbf {\bibinfo {volume} {60}},\ \bibinfo
  {pages} {114028} (\bibinfo {year} {1999})},\ \Eprint
  {https://arxiv.org/abs/hep-ph/9903292} {arXiv:hep-ph/9903292} \BibitemShut
  {NoStop}%
\bibitem [{\citenamefont {Qin}\ \emph {et~al.}(2011)\citenamefont {Qin},
  \citenamefont {Chang}, \citenamefont {Chen}, \citenamefont {Liu},\ and\
  \citenamefont {Roberts}}]{Qin:2010nq}%
  \BibitemOpen
  \bibfield  {author} {\bibinfo {author} {\bibfnamefont {S.-x.}\ \bibnamefont
  {Qin}}, \bibinfo {author} {\bibfnamefont {L.}~\bibnamefont {Chang}}, \bibinfo
  {author} {\bibfnamefont {H.}~\bibnamefont {Chen}}, \bibinfo {author}
  {\bibfnamefont {Y.-x.}\ \bibnamefont {Liu}},\ and\ \bibinfo {author}
  {\bibfnamefont {C.~D.}\ \bibnamefont {Roberts}},\ }\bibfield  {title}
  {\bibinfo {title} {{Phase diagram and critical endpoint for
  strongly-interacting quarks}},\ }\href
  {https://doi.org/10.1103/PhysRevLett.106.172301} {\bibfield  {journal}
  {\bibinfo  {journal} {Phys. Rev. Lett.}\ }\textbf {\bibinfo {volume} {106}},\
  \bibinfo {pages} {172301} (\bibinfo {year} {2011})},\ \Eprint
  {https://arxiv.org/abs/1011.2876} {arXiv:1011.2876 [nucl-th]} \BibitemShut
  {NoStop}%
\bibitem [{\citenamefont {Fu}\ \emph {et~al.}(2020)\citenamefont {Fu},
  \citenamefont {Pawlowski},\ and\ \citenamefont {Rennecke}}]{Fu:2019hdw}%
  \BibitemOpen
  \bibfield  {author} {\bibinfo {author} {\bibfnamefont {W.-j.}\ \bibnamefont
  {Fu}}, \bibinfo {author} {\bibfnamefont {J.~M.}\ \bibnamefont {Pawlowski}},\
  and\ \bibinfo {author} {\bibfnamefont {F.}~\bibnamefont {Rennecke}},\
  }\bibfield  {title} {\bibinfo {title} {{QCD phase structure at finite
  temperature and density}},\ }\href
  {https://doi.org/10.1103/PhysRevD.101.054032} {\bibfield  {journal} {\bibinfo
   {journal} {Phys. Rev. D}\ }\textbf {\bibinfo {volume} {101}},\ \bibinfo
  {pages} {054032} (\bibinfo {year} {2020})},\ \Eprint
  {https://arxiv.org/abs/1909.02991} {arXiv:1909.02991 [hep-ph]} \BibitemShut
  {NoStop}%
\bibitem [{\citenamefont {Gao}\ and\ \citenamefont
  {Pawlowski}(2021)}]{Gao:2020fbl}%
  \BibitemOpen
  \bibfield  {author} {\bibinfo {author} {\bibfnamefont {F.}~\bibnamefont
  {Gao}}\ and\ \bibinfo {author} {\bibfnamefont {J.~M.}\ \bibnamefont
  {Pawlowski}},\ }\bibfield  {title} {\bibinfo {title} {{Chiral phase structure
  and critical end point in QCD}},\ }\href
  {https://doi.org/10.1016/j.physletb.2021.136584} {\bibfield  {journal}
  {\bibinfo  {journal} {Phys. Lett. B}\ }\textbf {\bibinfo {volume} {820}},\
  \bibinfo {pages} {136584} (\bibinfo {year} {2021})},\ \Eprint
  {https://arxiv.org/abs/2010.13705} {arXiv:2010.13705 [hep-ph]} \BibitemShut
  {NoStop}%
\bibitem [{\citenamefont {Critelli}\ \emph {et~al.}(2017)\citenamefont
  {Critelli}, \citenamefont {Noronha}, \citenamefont {Noronha-Hostler},
  \citenamefont {Portillo}, \citenamefont {Ratti},\ and\ \citenamefont
  {Rougemont}}]{Critelli:2017ouy}%
  \BibitemOpen
  \bibfield  {author} {\bibinfo {author} {\bibfnamefont {R.}~\bibnamefont
  {Critelli}}, \bibinfo {author} {\bibfnamefont {J.}~\bibnamefont {Noronha}},
  \bibinfo {author} {\bibfnamefont {J.}~\bibnamefont {Noronha-Hostler}},
  \bibinfo {author} {\bibfnamefont {I.}~\bibnamefont {Portillo}}, \bibinfo
  {author} {\bibfnamefont {C.}~\bibnamefont {Ratti}},\ and\ \bibinfo {author}
  {\bibfnamefont {R.}~\bibnamefont {Rougemont}},\ }\bibfield  {title} {\bibinfo
  {title} {{Critical point in the phase diagram of primordial quark-gluon
  matter from black hole physics}},\ }\href
  {https://doi.org/10.1103/PhysRevD.96.096026} {\bibfield  {journal} {\bibinfo
  {journal} {Phys. Rev. D}\ }\textbf {\bibinfo {volume} {96}},\ \bibinfo
  {pages} {096026} (\bibinfo {year} {2017})},\ \Eprint
  {https://arxiv.org/abs/1706.00455} {arXiv:1706.00455 [nucl-th]} \BibitemShut
  {NoStop}%
\bibitem [{\citenamefont {Cai}\ \emph {et~al.}(2022)\citenamefont {Cai},
  \citenamefont {He}, \citenamefont {Li},\ and\ \citenamefont
  {Wang}}]{Cai:2022omk}%
  \BibitemOpen
  \bibfield  {author} {\bibinfo {author} {\bibfnamefont {R.-G.}\ \bibnamefont
  {Cai}}, \bibinfo {author} {\bibfnamefont {S.}~\bibnamefont {He}}, \bibinfo
  {author} {\bibfnamefont {L.}~\bibnamefont {Li}},\ and\ \bibinfo {author}
  {\bibfnamefont {Y.-X.}\ \bibnamefont {Wang}},\ }\bibfield  {title} {\bibinfo
  {title} {{Probing QCD critical point and induced gravitational wave by black
  hole physics}},\ }\href {https://doi.org/10.1103/PhysRevD.106.L121902}
  {\bibfield  {journal} {\bibinfo  {journal} {Phys. Rev. D}\ }\textbf {\bibinfo
  {volume} {106}},\ \bibinfo {pages} {L121902} (\bibinfo {year} {2022})},\
  \Eprint {https://arxiv.org/abs/2201.02004} {arXiv:2201.02004 [hep-th]}
  \BibitemShut {NoStop}%
\bibitem [{\citenamefont {Li}\ \emph {et~al.}(2026)\citenamefont {Li},
  \citenamefont {Li},\ and\ \citenamefont {Huang}}]{Li:2026lxx}%
  \BibitemOpen
  \bibfield  {author} {\bibinfo {author} {\bibfnamefont {Z.}~\bibnamefont
  {Li}}, \bibinfo {author} {\bibfnamefont {D.}~\bibnamefont {Li}},\ and\
  \bibinfo {author} {\bibfnamefont {M.}~\bibnamefont {Huang}},\ }\href@noop {}
  {\bibinfo {title} {{Flavor-Dependent QCD Critical Endpoint and Dual-Channel
  Fluctuations from Multi-Charge Holography}}} (\bibinfo {year} {2026}),\
  \Eprint {https://arxiv.org/abs/2607.19149} {arXiv:2607.19149 [hep-ph]}
  \BibitemShut {NoStop}%
\bibitem [{\citenamefont {Borsanyi}\ \emph {et~al.}(2025)\citenamefont
  {Borsanyi}, \citenamefont {Fodor}, \citenamefont {Guenther}, \citenamefont
  {Parotto}, \citenamefont {Pasztor}, \citenamefont {Ratti}, \citenamefont
  {Vovchenko},\ and\ \citenamefont {Wong}}]{Borsanyi:2025dyp}%
  \BibitemOpen
  \bibfield  {author} {\bibinfo {author} {\bibfnamefont {S.}~\bibnamefont
  {Borsanyi}}, \bibinfo {author} {\bibfnamefont {Z.}~\bibnamefont {Fodor}},
  \bibinfo {author} {\bibfnamefont {J.~N.}\ \bibnamefont {Guenther}}, \bibinfo
  {author} {\bibfnamefont {P.}~\bibnamefont {Parotto}}, \bibinfo {author}
  {\bibfnamefont {A.}~\bibnamefont {Pasztor}}, \bibinfo {author} {\bibfnamefont
  {C.}~\bibnamefont {Ratti}}, \bibinfo {author} {\bibfnamefont
  {V.}~\bibnamefont {Vovchenko}},\ and\ \bibinfo {author} {\bibfnamefont
  {C.~H.}\ \bibnamefont {Wong}},\ }\bibfield  {title} {\bibinfo {title}
  {{Lattice QCD constraints on the critical point from an improved precision
  equation of state}},\ }\href {https://doi.org/10.1103/rj6r-dmg9} {\bibfield
  {journal} {\bibinfo  {journal} {Phys. Rev. D}\ }\textbf {\bibinfo {volume}
  {112}},\ \bibinfo {pages} {L111505} (\bibinfo {year} {2025})},\ \Eprint
  {https://arxiv.org/abs/2502.10267} {arXiv:2502.10267 [hep-lat]} \BibitemShut
  {NoStop}%
\bibitem [{\citenamefont {Clarke}\ \emph {et~al.}(2025)\citenamefont {Clarke},
  \citenamefont {Dimopoulos}, \citenamefont {Di~Renzo}, \citenamefont
  {Goswami}, \citenamefont {Schmidt}, \citenamefont {Singh},\ and\
  \citenamefont {Zambello}}]{Clarke:2024ugt}%
  \BibitemOpen
  \bibfield  {author} {\bibinfo {author} {\bibfnamefont {D.~A.}\ \bibnamefont
  {Clarke}}, \bibinfo {author} {\bibfnamefont {P.}~\bibnamefont {Dimopoulos}},
  \bibinfo {author} {\bibfnamefont {F.}~\bibnamefont {Di~Renzo}}, \bibinfo
  {author} {\bibfnamefont {J.}~\bibnamefont {Goswami}}, \bibinfo {author}
  {\bibfnamefont {C.}~\bibnamefont {Schmidt}}, \bibinfo {author} {\bibfnamefont
  {S.}~\bibnamefont {Singh}},\ and\ \bibinfo {author} {\bibfnamefont
  {K.}~\bibnamefont {Zambello}},\ }\bibfield  {title} {\bibinfo {title}
  {{Searching for the QCD critical end point using multipoint Pad{\'e}
  approximations}},\ }\href {https://doi.org/10.1103/y6kg-ry8x} {\bibfield
  {journal} {\bibinfo  {journal} {Phys. Rev. D}\ }\textbf {\bibinfo {volume}
  {112}},\ \bibinfo {pages} {L091504} (\bibinfo {year} {2025})},\ \Eprint
  {https://arxiv.org/abs/2405.10196} {arXiv:2405.10196 [hep-lat]} \BibitemShut
  {NoStop}%
\bibitem [{\citenamefont {Adamczyk}\ \emph {et~al.}(2017)\citenamefont
  {Adamczyk} \emph {et~al.}}]{Adamczyk:2017iwn}%
  \BibitemOpen
  \bibfield  {author} {\bibinfo {author} {\bibfnamefont {L.}~\bibnamefont
  {Adamczyk}} \emph {et~al.} (\bibinfo {collaboration} {STAR}),\ }\bibfield
  {title} {\bibinfo {title} {{Bulk Properties of the Medium Produced in
  Relativistic Heavy-Ion Collisions from the Beam Energy Scan Program}},\
  }\href {https://doi.org/10.1103/PhysRevC.96.044904} {\bibfield  {journal}
  {\bibinfo  {journal} {Phys. Rev. C}\ }\textbf {\bibinfo {volume} {96}},\
  \bibinfo {pages} {044904} (\bibinfo {year} {2017})},\ \Eprint
  {https://arxiv.org/abs/1701.07065} {arXiv:1701.07065 [nucl-ex]} \BibitemShut
  {NoStop}%
\bibitem [{\citenamefont {Luo}\ and\ \citenamefont {Xu}(2017)}]{Luo:2017faz}%
  \BibitemOpen
  \bibfield  {author} {\bibinfo {author} {\bibfnamefont {X.}~\bibnamefont
  {Luo}}\ and\ \bibinfo {author} {\bibfnamefont {N.}~\bibnamefont {Xu}},\
  }\bibfield  {title} {\bibinfo {title} {{Search for the QCD Critical Point
  with Fluctuations of Conserved Quantities in Relativistic Heavy-Ion
  Collisions at RHIC : An Overview}},\ }\href
  {https://doi.org/10.1007/s41365-017-0257-0} {\bibfield  {journal} {\bibinfo
  {journal} {Nucl. Sci. Tech.}\ }\textbf {\bibinfo {volume} {28}},\ \bibinfo
  {pages} {112} (\bibinfo {year} {2017})},\ \Eprint
  {https://arxiv.org/abs/1701.02105} {arXiv:1701.02105 [nucl-ex]} \BibitemShut
  {NoStop}%
\bibitem [{\citenamefont {Adam}\ \emph {et~al.}(2021)\citenamefont {Adam} \emph
  {et~al.}}]{STAR:2020tga}%
  \BibitemOpen
  \bibfield  {author} {\bibinfo {author} {\bibfnamefont {J.}~\bibnamefont
  {Adam}} \emph {et~al.} (\bibinfo {collaboration} {STAR}),\ }\bibfield
  {title} {\bibinfo {title} {{Nonmonotonic Energy Dependence of Net-Proton
  Number Fluctuations}},\ }\href
  {https://doi.org/10.1103/PhysRevLett.134.139902} {\bibfield  {journal}
  {\bibinfo  {journal} {Phys. Rev. Lett.}\ }\textbf {\bibinfo {volume} {126}},\
  \bibinfo {pages} {092301} (\bibinfo {year} {2021})},\ \bibinfo {note}
  {[Erratum: Phys.Rev.Lett. 134, 139902 (2025)]},\ \Eprint
  {https://arxiv.org/abs/2001.02852} {arXiv:2001.02852 [nucl-ex]} \BibitemShut
  {NoStop}%
\bibitem [{\citenamefont {Aboona}\ \emph {et~al.}(2025)\citenamefont {Aboona}
  \emph {et~al.}}]{STAR:2025zdq}%
  \BibitemOpen
  \bibfield  {author} {\bibinfo {author} {\bibfnamefont {B.~E.}\ \bibnamefont
  {Aboona}} \emph {et~al.} (\bibinfo {collaboration} {STAR}),\ }\bibfield
  {title} {\bibinfo {title} {{Precision Measurement of Net-Proton-Number
  Fluctuations in Au+Au Collisions at RHIC}},\ }\href
  {https://doi.org/10.1103/9l69-2d7p} {\bibfield  {journal} {\bibinfo
  {journal} {Phys. Rev. Lett.}\ }\textbf {\bibinfo {volume} {135}},\ \bibinfo
  {pages} {142301} (\bibinfo {year} {2025})},\ \Eprint
  {https://arxiv.org/abs/2504.00817} {arXiv:2504.00817 [nucl-ex]} \BibitemShut
  {NoStop}%
\bibitem [{\citenamefont {Zhou}\ and\ \citenamefont
  {Yang}(2022)}]{Zhou:2022pxl}%
  \BibitemOpen
  \bibfield  {author} {\bibinfo {author} {\bibfnamefont {X.}~\bibnamefont
  {Zhou}}\ and\ \bibinfo {author} {\bibfnamefont {J.}~\bibnamefont {Yang}}
  (\bibinfo {collaboration} {HIAF project Team}),\ }\bibfield  {title}
  {\bibinfo {title} {{Status of the high-intensity heavy-ion accelerator
  facility in China}},\ }\href {https://doi.org/10.1007/s43673-022-00064-1}
  {\bibfield  {journal} {\bibinfo  {journal} {AAPPS Bull.}\ }\textbf {\bibinfo
  {volume} {32}},\ \bibinfo {pages} {35} (\bibinfo {year} {2022})}\BibitemShut
  {NoStop}%
\bibitem [{\citenamefont {Ablyazimov}\ \emph {et~al.}(2017)\citenamefont
  {Ablyazimov} \emph {et~al.}}]{CBM:2016kpk}%
  \BibitemOpen
  \bibfield  {author} {\bibinfo {author} {\bibfnamefont {T.}~\bibnamefont
  {Ablyazimov}} \emph {et~al.} (\bibinfo {collaboration} {CBM}),\ }\bibfield
  {title} {\bibinfo {title} {{Challenges in QCD matter physics --The scientific
  programme of the Compressed Baryonic Matter experiment at FAIR}},\ }\href
  {https://doi.org/10.1140/epja/i2017-12248-y} {\bibfield  {journal} {\bibinfo
  {journal} {Eur. Phys. J. A}\ }\textbf {\bibinfo {volume} {53}},\ \bibinfo
  {pages} {60} (\bibinfo {year} {2017})},\ \Eprint
  {https://arxiv.org/abs/1607.01487} {arXiv:1607.01487 [nucl-ex]} \BibitemShut
  {NoStop}%
\bibitem [{\citenamefont {Abgaryan}\ \emph {et~al.}(2022)\citenamefont
  {Abgaryan} \emph {et~al.}}]{MPD:2022qhn}%
  \BibitemOpen
  \bibfield  {author} {\bibinfo {author} {\bibfnamefont {V.}~\bibnamefont
  {Abgaryan}} \emph {et~al.} (\bibinfo {collaboration} {MPD}),\ }\bibfield
  {title} {\bibinfo {title} {{Status and initial physics performance studies of
  the MPD experiment at NICA}},\ }\href
  {https://doi.org/10.1140/epja/s10050-022-00750-6} {\bibfield  {journal}
  {\bibinfo  {journal} {Eur. Phys. J. A}\ }\textbf {\bibinfo {volume} {58}},\
  \bibinfo {pages} {140} (\bibinfo {year} {2022})},\ \Eprint
  {https://arxiv.org/abs/2202.08970} {arXiv:2202.08970 [physics.ins-det]}
  \BibitemShut {NoStop}%
\bibitem [{\citenamefont {Bzdak}\ \emph {et~al.}(2020)\citenamefont {Bzdak},
  \citenamefont {Esumi}, \citenamefont {Koch}, \citenamefont {Liao},
  \citenamefont {Stephanov},\ and\ \citenamefont {Xu}}]{Bzdak:2019pkr}%
  \BibitemOpen
  \bibfield  {author} {\bibinfo {author} {\bibfnamefont {A.}~\bibnamefont
  {Bzdak}}, \bibinfo {author} {\bibfnamefont {S.}~\bibnamefont {Esumi}},
  \bibinfo {author} {\bibfnamefont {V.}~\bibnamefont {Koch}}, \bibinfo {author}
  {\bibfnamefont {J.}~\bibnamefont {Liao}}, \bibinfo {author} {\bibfnamefont
  {M.}~\bibnamefont {Stephanov}},\ and\ \bibinfo {author} {\bibfnamefont
  {N.}~\bibnamefont {Xu}},\ }\bibfield  {title} {\bibinfo {title} {{Mapping the
  Phases of Quantum Chromodynamics with Beam Energy Scan}},\ }\href
  {https://doi.org/10.1016/j.physrep.2020.01.005} {\bibfield  {journal}
  {\bibinfo  {journal} {Phys. Rept.}\ }\textbf {\bibinfo {volume} {853}},\
  \bibinfo {pages} {1} (\bibinfo {year} {2020})},\ \Eprint
  {https://arxiv.org/abs/1906.00936} {arXiv:1906.00936 [nucl-th]} \BibitemShut
  {NoStop}%
\bibitem [{\citenamefont {Abdallah}\ \emph {et~al.}(2022)\citenamefont
  {Abdallah} \emph {et~al.}}]{STAR:2021fge}%
  \BibitemOpen
  \bibfield  {author} {\bibinfo {author} {\bibfnamefont {M.~S.}\ \bibnamefont
  {Abdallah}} \emph {et~al.} (\bibinfo {collaboration} {STAR}),\ }\bibfield
  {title} {\bibinfo {title} {{Measurements of Proton High Order Cumulants in
  $\sqrt{s_{_{\mathrm{NN}}}}$ = 3 GeV Au+Au Collisions and Implications for the
  QCD Critical Point}},\ }\href
  {https://doi.org/10.1103/PhysRevLett.128.202303} {\bibfield  {journal}
  {\bibinfo  {journal} {Phys. Rev. Lett.}\ }\textbf {\bibinfo {volume} {128}},\
  \bibinfo {pages} {202303} (\bibinfo {year} {2022})},\ \Eprint
  {https://arxiv.org/abs/2112.00240} {arXiv:2112.00240 [nucl-ex]} \BibitemShut
  {NoStop}%
\bibitem [{\citenamefont {Li}\ \emph {et~al.}(2018)\citenamefont {Li},
  \citenamefont {Chen}, \citenamefont {Li},\ and\ \citenamefont
  {Huang}}]{Li:2017ple}%
  \BibitemOpen
  \bibfield  {author} {\bibinfo {author} {\bibfnamefont {Z.}~\bibnamefont
  {Li}}, \bibinfo {author} {\bibfnamefont {Y.}~\bibnamefont {Chen}}, \bibinfo
  {author} {\bibfnamefont {D.}~\bibnamefont {Li}},\ and\ \bibinfo {author}
  {\bibfnamefont {M.}~\bibnamefont {Huang}},\ }\bibfield  {title} {\bibinfo
  {title} {{Locating the QCD critical end point through the peaked baryon
  number susceptibilities along the freeze-out line}},\ }\href
  {https://doi.org/10.1088/1674-1137/42/1/013103} {\bibfield  {journal}
  {\bibinfo  {journal} {Chin. Phys. C}\ }\textbf {\bibinfo {volume} {42}},\
  \bibinfo {pages} {013103} (\bibinfo {year} {2018})},\ \Eprint
  {https://arxiv.org/abs/1706.02238} {arXiv:1706.02238 [hep-ph]} \BibitemShut
  {NoStop}%
\bibitem [{\citenamefont {Stodolsky}(1995)}]{Stodolsky:1995ds}%
  \BibitemOpen
  \bibfield  {author} {\bibinfo {author} {\bibfnamefont {L.}~\bibnamefont
  {Stodolsky}},\ }\bibfield  {title} {\bibinfo {title} {{Temperature
  fluctuations in multiparticle production}},\ }\href
  {https://doi.org/10.1103/PhysRevLett.75.1044} {\bibfield  {journal} {\bibinfo
   {journal} {Phys. Rev. Lett.}\ }\textbf {\bibinfo {volume} {75}},\ \bibinfo
  {pages} {1044} (\bibinfo {year} {1995})}\BibitemShut {NoStop}%
\bibitem [{\citenamefont {Shuryak}(1998)}]{Shuryak:1997yj}%
  \BibitemOpen
  \bibfield  {author} {\bibinfo {author} {\bibfnamefont {E.~V.}\ \bibnamefont
  {Shuryak}},\ }\bibfield  {title} {\bibinfo {title} {{Event per event analysis
  of heavy ion collisions and thermodynamical fluctuations}},\ }\href
  {https://doi.org/10.1016/S0370-2693(98)00127-0} {\bibfield  {journal}
  {\bibinfo  {journal} {Phys. Lett. B}\ }\textbf {\bibinfo {volume} {423}},\
  \bibinfo {pages} {9} (\bibinfo {year} {1998})},\ \Eprint
  {https://arxiv.org/abs/hep-ph/9704456} {arXiv:hep-ph/9704456} \BibitemShut
  {NoStop}%
\bibitem [{\citenamefont {Gavin}(2004)}]{Gavin:2003cb}%
  \BibitemOpen
  \bibfield  {author} {\bibinfo {author} {\bibfnamefont {S.}~\bibnamefont
  {Gavin}},\ }\bibfield  {title} {\bibinfo {title} {{Traces of thermalization
  from transverse momentum fluctuations in nuclear collisions}},\ }\href
  {https://doi.org/10.1103/PhysRevLett.92.162301} {\bibfield  {journal}
  {\bibinfo  {journal} {Phys. Rev. Lett.}\ }\textbf {\bibinfo {volume} {92}},\
  \bibinfo {pages} {162301} (\bibinfo {year} {2004})},\ \Eprint
  {https://arxiv.org/abs/nucl-th/0308067} {arXiv:nucl-th/0308067} \BibitemShut
  {NoStop}%
\bibitem [{\citenamefont {Gardim}\ \emph {et~al.}(2020)\citenamefont {Gardim},
  \citenamefont {Giacalone}, \citenamefont {Luzum},\ and\ \citenamefont
  {Ollitrault}}]{Gardim:2019xjs}%
  \BibitemOpen
  \bibfield  {author} {\bibinfo {author} {\bibfnamefont {F.~G.}\ \bibnamefont
  {Gardim}}, \bibinfo {author} {\bibfnamefont {G.}~\bibnamefont {Giacalone}},
  \bibinfo {author} {\bibfnamefont {M.}~\bibnamefont {Luzum}},\ and\ \bibinfo
  {author} {\bibfnamefont {J.-Y.}\ \bibnamefont {Ollitrault}},\ }\bibfield
  {title} {\bibinfo {title} {{Thermodynamics of hot strong-interaction matter
  from ultrarelativistic nuclear collisions}},\ }\href
  {https://doi.org/10.1038/s41567-020-0846-4} {\bibfield  {journal} {\bibinfo
  {journal} {Nature Phys.}\ }\textbf {\bibinfo {volume} {16}},\ \bibinfo
  {pages} {615} (\bibinfo {year} {2020})},\ \Eprint
  {https://arxiv.org/abs/1908.09728} {arXiv:1908.09728 [nucl-th]} \BibitemShut
  {NoStop}%
\bibitem [{\citenamefont {Giacalone}\ \emph {et~al.}(2021)\citenamefont
  {Giacalone}, \citenamefont {Gardim}, \citenamefont {Noronha-Hostler},\ and\
  \citenamefont {Ollitrault}}]{Giacalone:2020lbm}%
  \BibitemOpen
  \bibfield  {author} {\bibinfo {author} {\bibfnamefont {G.}~\bibnamefont
  {Giacalone}}, \bibinfo {author} {\bibfnamefont {F.~G.}\ \bibnamefont
  {Gardim}}, \bibinfo {author} {\bibfnamefont {J.}~\bibnamefont
  {Noronha-Hostler}},\ and\ \bibinfo {author} {\bibfnamefont {J.-Y.}\
  \bibnamefont {Ollitrault}},\ }\bibfield  {title} {\bibinfo {title} {{Skewness
  of mean transverse momentum fluctuations in heavy-ion collisions}},\ }\href
  {https://doi.org/10.1103/PhysRevC.103.024910} {\bibfield  {journal} {\bibinfo
   {journal} {Phys. Rev. C}\ }\textbf {\bibinfo {volume} {103}},\ \bibinfo
  {pages} {024910} (\bibinfo {year} {2021})},\ \Eprint
  {https://arxiv.org/abs/2004.09799} {arXiv:2004.09799 [nucl-th]} \BibitemShut
  {NoStop}%
\bibitem [{\citenamefont {Acharya}\ \emph {et~al.}(2024)\citenamefont {Acharya}
  \emph {et~al.}}]{ALICE:2023tej}%
  \BibitemOpen
  \bibfield  {author} {\bibinfo {author} {\bibfnamefont {S.}~\bibnamefont
  {Acharya}} \emph {et~al.} (\bibinfo {collaboration} {ALICE}),\ }\bibfield
  {title} {\bibinfo {title} {{Skewness and kurtosis of mean transverse momentum
  fluctuations at the LHC energies}},\ }\href
  {https://doi.org/10.1016/j.physletb.2024.138541} {\bibfield  {journal}
  {\bibinfo  {journal} {Phys. Lett. B}\ }\textbf {\bibinfo {volume} {850}},\
  \bibinfo {pages} {138541} (\bibinfo {year} {2024})},\ \Eprint
  {https://arxiv.org/abs/2308.16217} {arXiv:2308.16217 [nucl-ex]} \BibitemShut
  {NoStop}%
\bibitem [{\citenamefont {Aad}\ \emph {et~al.}(2024)\citenamefont {Aad} \emph
  {et~al.}}]{ATLAS:2024jvf}%
  \BibitemOpen
  \bibfield  {author} {\bibinfo {author} {\bibfnamefont {G.}~\bibnamefont
  {Aad}} \emph {et~al.} (\bibinfo {collaboration} {ATLAS}),\ }\bibfield
  {title} {\bibinfo {title} {{Disentangling Sources of Momentum Fluctuations in
  Xe+Xe and Pb+Pb Collisions with the ATLAS Detector}},\ }\href
  {https://doi.org/10.1103/PhysRevLett.133.252301} {\bibfield  {journal}
  {\bibinfo  {journal} {Phys. Rev. Lett.}\ }\textbf {\bibinfo {volume} {133}},\
  \bibinfo {pages} {252301} (\bibinfo {year} {2024})},\ \Eprint
  {https://arxiv.org/abs/2407.06413} {arXiv:2407.06413 [nucl-ex]} \BibitemShut
  {NoStop}%
\bibitem [{\citenamefont {Aboona}\ \emph {et~al.}(2026)\citenamefont {Aboona}
  \emph {et~al.}}]{STAR:2026vjv}%
  \BibitemOpen
  \bibfield  {author} {\bibinfo {author} {\bibfnamefont {B.~E.}\ \bibnamefont
  {Aboona}} \emph {et~al.} (\bibinfo {collaboration} {STAR}),\ }\bibfield
  {title} {\bibinfo {title} {{Nonmonotonicity of Transverse Momentum
  Correlations in Au+Au Collisions at RHIC}},\ }\href
  {https://doi.org/10.1103/2xsn-rgx3} {\bibfield  {journal} {\bibinfo
  {journal} {Phys. Rev. Lett.}\ }\textbf {\bibinfo {volume} {137}},\ \bibinfo
  {pages} {132301} (\bibinfo {year} {2026})},\ \Eprint
  {https://arxiv.org/abs/2604.06434} {arXiv:2604.06434 [nucl-ex]} \BibitemShut
  {NoStop}%
\bibitem [{\citenamefont {Reichert}\ \emph {et~al.}(2026)\citenamefont
  {Reichert}, \citenamefont {Steinheimer},\ and\ \citenamefont
  {Bleicher}}]{Reichert:2026ztj}%
  \BibitemOpen
  \bibfield  {author} {\bibinfo {author} {\bibfnamefont {T.}~\bibnamefont
  {Reichert}}, \bibinfo {author} {\bibfnamefont {J.}~\bibnamefont
  {Steinheimer}},\ and\ \bibinfo {author} {\bibfnamefont {M.}~\bibnamefont
  {Bleicher}},\ }\href@noop {} {\bibinfo {title} {{Non-monotonicity of $p_T$
  correlations from meson-baryon mixing}}} (\bibinfo {year} {2026}),\ \Eprint
  {https://arxiv.org/abs/2606.03294} {arXiv:2606.03294 [nucl-th]} \BibitemShut
  {NoStop}%
\bibitem [{\citenamefont {Liu}\ \emph {et~al.}(2026{\natexlab{a}})\citenamefont
  {Liu}, \citenamefont {Zhang}, \citenamefont {Zhang}, \citenamefont {Chen},\
  and\ \citenamefont {Ma}}]{Liu:2026lnr}%
  \BibitemOpen
  \bibfield  {author} {\bibinfo {author} {\bibfnamefont {Y.}~\bibnamefont
  {Liu}}, \bibinfo {author} {\bibfnamefont {L.}~\bibnamefont {Zhang}}, \bibinfo
  {author} {\bibfnamefont {C.}~\bibnamefont {Zhang}}, \bibinfo {author}
  {\bibfnamefont {J.}~\bibnamefont {Chen}},\ and\ \bibinfo {author}
  {\bibfnamefont {C.}~\bibnamefont {Ma}},\ }\href@noop {} {\bibinfo {title}
  {{Mean-$p_T$ fluctuations in Au+Au collisions at $\sqrt{s_{\rm
  NN}}=3.0$--$19.6$ GeV within JAM2}}} (\bibinfo {year} {2026}{\natexlab{a}}),\
  \Eprint {https://arxiv.org/abs/2609.28074} {arXiv:2609.28074 [nucl-ex]}
  \BibitemShut {NoStop}%
\bibitem [{\citenamefont {Chen}\ \emph {et~al.}(2026)\citenamefont {Chen},
  \citenamefont {Fu}, \citenamefont {Yin},\ and\ \citenamefont
  {Zhang}}]{Chen:2025vwl}%
  \BibitemOpen
  \bibfield  {author} {\bibinfo {author} {\bibfnamefont {J.}~\bibnamefont
  {Chen}}, \bibinfo {author} {\bibfnamefont {W.-j.}\ \bibnamefont {Fu}},
  \bibinfo {author} {\bibfnamefont {S.}~\bibnamefont {Yin}},\ and\ \bibinfo
  {author} {\bibfnamefont {C.}~\bibnamefont {Zhang}},\ }\bibfield  {title}
  {\bibinfo {title} {{High-order fluctuations of temperature in hot QCD
  matter}},\ }\href {https://doi.org/10.1103/919p-wjfl} {\bibfield  {journal}
  {\bibinfo  {journal} {Phys. Rev. D}\ }\textbf {\bibinfo {volume} {114}},\
  \bibinfo {pages} {014064} (\bibinfo {year} {2026})},\ \Eprint
  {https://arxiv.org/abs/2504.06886} {arXiv:2504.06886 [hep-ph]} \BibitemShut
  {NoStop}%
\bibitem [{\citenamefont {Liu}\ \emph {et~al.}(2026{\natexlab{b}})\citenamefont
  {Liu}, \citenamefont {Wu}, \citenamefont {Liu},\ and\ \citenamefont
  {Chu}}]{Liu:2025cpj}%
  \BibitemOpen
  \bibfield  {author} {\bibinfo {author} {\bibfnamefont {H.}~\bibnamefont
  {Liu}}, \bibinfo {author} {\bibfnamefont {P.}~\bibnamefont {Wu}}, \bibinfo
  {author} {\bibfnamefont {H.-M.}\ \bibnamefont {Liu}},\ and\ \bibinfo {author}
  {\bibfnamefont {P.-C.}\ \bibnamefont {Chu}},\ }\bibfield  {title} {\bibinfo
  {title} {{Fluctuations of Temperature in the Polyakov Loop-Extended
  Nambu{\textendash}Jona-Lasinio Model}},\ }\href
  {https://doi.org/10.3390/universe12020037} {\bibfield  {journal} {\bibinfo
  {journal} {Universe}\ }\textbf {\bibinfo {volume} {12}},\ \bibinfo {pages}
  {37} (\bibinfo {year} {2026}{\natexlab{b}})},\ \Eprint
  {https://arxiv.org/abs/2512.23934} {arXiv:2512.23934 [nucl-th]} \BibitemShut
  {NoStop}%
\bibitem [{\citenamefont {Liu}\ \emph {et~al.}(2026{\natexlab{c}})\citenamefont
  {Liu}, \citenamefont {Wu}, \citenamefont {Liu},\ and\ \citenamefont
  {Chu}}]{Liu:2026pro}%
  \BibitemOpen
  \bibfield  {author} {\bibinfo {author} {\bibfnamefont {H.}~\bibnamefont
  {Liu}}, \bibinfo {author} {\bibfnamefont {P.}~\bibnamefont {Wu}}, \bibinfo
  {author} {\bibfnamefont {H.-M.}\ \bibnamefont {Liu}},\ and\ \bibinfo {author}
  {\bibfnamefont {P.-C.}\ \bibnamefont {Chu}},\ }\href@noop {} {\bibinfo
  {title} {{Temperature fluctuations in a realistic Polyakov-loop extended
  Nambu--Jona-Lasinio Model along the freeze-out line}}} (\bibinfo {year}
  {2026}{\natexlab{c}}),\ \Eprint {https://arxiv.org/abs/2607.19275}
  {arXiv:2607.19275 [nucl-th]} \BibitemShut {NoStop}%
\bibitem [{\citenamefont {Mallick}(2026)}]{Mallick:2026dsc}%
  \BibitemOpen
  \bibfield  {author} {\bibinfo {author} {\bibfnamefont {D.}~\bibnamefont
  {Mallick}},\ }\href@noop {} {\bibinfo {title} {{Interaction fingerprints in
  temperature-fluctuation cumulants from the QCD crossover to nuclear
  liquid-gas criticality}}} (\bibinfo {year} {2026}),\ \Eprint
  {https://arxiv.org/abs/2607.24616} {arXiv:2607.24616 [nucl-th]} \BibitemShut
  {NoStop}%
\bibitem [{\citenamefont {Gursoy}\ and\ \citenamefont
  {Kiritsis}(2008)}]{Gursoy:2007cb}%
  \BibitemOpen
  \bibfield  {author} {\bibinfo {author} {\bibfnamefont {U.}~\bibnamefont
  {Gursoy}}\ and\ \bibinfo {author} {\bibfnamefont {E.}~\bibnamefont
  {Kiritsis}},\ }\bibfield  {title} {\bibinfo {title} {{Exploring improved
  holographic theories for QCD: Part I}},\ }\href
  {https://doi.org/10.1088/1126-6708/2008/02/032} {\bibfield  {journal}
  {\bibinfo  {journal} {JHEP}\ }\textbf {\bibinfo {volume} {02}},\ \bibinfo
  {pages} {032}},\ \Eprint {https://arxiv.org/abs/0707.1324} {arXiv:0707.1324
  [hep-th]} \BibitemShut {NoStop}%
\bibitem [{\citenamefont {Gursoy}\ \emph {et~al.}(2008)\citenamefont {Gursoy},
  \citenamefont {Kiritsis},\ and\ \citenamefont {Nitti}}]{Gursoy:2007er}%
  \BibitemOpen
  \bibfield  {author} {\bibinfo {author} {\bibfnamefont {U.}~\bibnamefont
  {Gursoy}}, \bibinfo {author} {\bibfnamefont {E.}~\bibnamefont {Kiritsis}},\
  and\ \bibinfo {author} {\bibfnamefont {F.}~\bibnamefont {Nitti}},\ }\bibfield
   {title} {\bibinfo {title} {{Exploring improved holographic theories for QCD:
  Part II}},\ }\href {https://doi.org/10.1088/1126-6708/2008/02/019} {\bibfield
   {journal} {\bibinfo  {journal} {JHEP}\ }\textbf {\bibinfo {volume} {02}},\
  \bibinfo {pages} {019}},\ \Eprint {https://arxiv.org/abs/0707.1349}
  {arXiv:0707.1349 [hep-th]} \BibitemShut {NoStop}%
\bibitem [{\citenamefont {Gursoy}\ \emph {et~al.}(2009)\citenamefont {Gursoy},
  \citenamefont {Kiritsis}, \citenamefont {Mazzanti},\ and\ \citenamefont
  {Nitti}}]{Gursoy:2009jd}%
  \BibitemOpen
  \bibfield  {author} {\bibinfo {author} {\bibfnamefont {U.}~\bibnamefont
  {Gursoy}}, \bibinfo {author} {\bibfnamefont {E.}~\bibnamefont {Kiritsis}},
  \bibinfo {author} {\bibfnamefont {L.}~\bibnamefont {Mazzanti}},\ and\
  \bibinfo {author} {\bibfnamefont {F.}~\bibnamefont {Nitti}},\ }\bibfield
  {title} {\bibinfo {title} {{Improved Holographic Yang-Mills at Finite
  Temperature: Comparison with Data}},\ }\href
  {https://doi.org/10.1016/j.nuclphysb.2009.05.017} {\bibfield  {journal}
  {\bibinfo  {journal} {Nucl. Phys. B}\ }\textbf {\bibinfo {volume} {820}},\
  \bibinfo {pages} {148} (\bibinfo {year} {2009})},\ \Eprint
  {https://arxiv.org/abs/0903.2859} {arXiv:0903.2859 [hep-th]} \BibitemShut
  {NoStop}%
\bibitem [{\citenamefont {Jarvinen}\ and\ \citenamefont
  {Kiritsis}(2012)}]{Jarvinen:2011qe}%
  \BibitemOpen
  \bibfield  {author} {\bibinfo {author} {\bibfnamefont {M.}~\bibnamefont
  {Jarvinen}}\ and\ \bibinfo {author} {\bibfnamefont {E.}~\bibnamefont
  {Kiritsis}},\ }\bibfield  {title} {\bibinfo {title} {{Holographic Models for
  QCD in the Veneziano Limit}},\ }\href
  {https://doi.org/10.1007/JHEP03(2012)002} {\bibfield  {journal} {\bibinfo
  {journal} {JHEP}\ }\textbf {\bibinfo {volume} {03}},\ \bibinfo {pages}
  {002}},\ \Eprint {https://arxiv.org/abs/1112.1261} {arXiv:1112.1261 [hep-ph]}
  \BibitemShut {NoStop}%
\bibitem [{\citenamefont {Alho}\ \emph {et~al.}(2014)\citenamefont {Alho},
  \citenamefont {J{\"a}rvinen}, \citenamefont {Kajantie}, \citenamefont
  {Kiritsis}, \citenamefont {Rosen},\ and\ \citenamefont
  {Tuominen}}]{Alho:2013hsa}%
  \BibitemOpen
  \bibfield  {author} {\bibinfo {author} {\bibfnamefont {T.}~\bibnamefont
  {Alho}}, \bibinfo {author} {\bibfnamefont {M.}~\bibnamefont {J{\"a}rvinen}},
  \bibinfo {author} {\bibfnamefont {K.}~\bibnamefont {Kajantie}}, \bibinfo
  {author} {\bibfnamefont {E.}~\bibnamefont {Kiritsis}}, \bibinfo {author}
  {\bibfnamefont {C.}~\bibnamefont {Rosen}},\ and\ \bibinfo {author}
  {\bibfnamefont {K.}~\bibnamefont {Tuominen}},\ }\bibfield  {title} {\bibinfo
  {title} {{A holographic model for QCD in the Veneziano limit at finite
  temperature and density}},\ }\href {https://doi.org/10.1007/JHEP04(2014)124}
  {\bibfield  {journal} {\bibinfo  {journal} {JHEP}\ }\textbf {\bibinfo
  {volume} {04}},\ \bibinfo {pages} {124}},\ \bibinfo {note} {[Erratum: JHEP
  02, 033 (2015)]},\ \Eprint {https://arxiv.org/abs/1312.5199} {arXiv:1312.5199
  [hep-ph]} \BibitemShut {NoStop}%
\bibitem [{\citenamefont {Li}\ and\ \citenamefont {Huang}(2013)}]{Li:2013oda}%
  \BibitemOpen
  \bibfield  {author} {\bibinfo {author} {\bibfnamefont {D.}~\bibnamefont
  {Li}}\ and\ \bibinfo {author} {\bibfnamefont {M.}~\bibnamefont {Huang}},\
  }\bibfield  {title} {\bibinfo {title} {{Dynamical holographic QCD model for
  glueball and light meson spectra}},\ }\href
  {https://doi.org/10.1007/JHEP11(2013)088} {\bibfield  {journal} {\bibinfo
  {journal} {JHEP}\ }\textbf {\bibinfo {volume} {11}},\ \bibinfo {pages}
  {088}},\ \Eprint {https://arxiv.org/abs/1303.6929} {arXiv:1303.6929 [hep-ph]}
  \BibitemShut {NoStop}%
\bibitem [{\citenamefont {Chelabi}\ \emph {et~al.}(2016)\citenamefont
  {Chelabi}, \citenamefont {Fang}, \citenamefont {Huang}, \citenamefont {Li},\
  and\ \citenamefont {Wu}}]{Chelabi:2015cwn}%
  \BibitemOpen
  \bibfield  {author} {\bibinfo {author} {\bibfnamefont {K.}~\bibnamefont
  {Chelabi}}, \bibinfo {author} {\bibfnamefont {Z.}~\bibnamefont {Fang}},
  \bibinfo {author} {\bibfnamefont {M.}~\bibnamefont {Huang}}, \bibinfo
  {author} {\bibfnamefont {D.}~\bibnamefont {Li}},\ and\ \bibinfo {author}
  {\bibfnamefont {Y.-L.}\ \bibnamefont {Wu}},\ }\bibfield  {title} {\bibinfo
  {title} {{Realization of chiral symmetry breaking and restoration in
  holographic QCD}},\ }\href {https://doi.org/10.1103/PhysRevD.93.101901}
  {\bibfield  {journal} {\bibinfo  {journal} {Phys. Rev. D}\ }\textbf {\bibinfo
  {volume} {93}},\ \bibinfo {pages} {101901} (\bibinfo {year} {2016})},\
  \Eprint {https://arxiv.org/abs/1511.02721} {arXiv:1511.02721 [hep-ph]}
  \BibitemShut {NoStop}%
\bibitem [{\citenamefont {Gubser}\ and\ \citenamefont
  {Nellore}(2008)}]{Gubser:2008ny}%
  \BibitemOpen
  \bibfield  {author} {\bibinfo {author} {\bibfnamefont {S.~S.}\ \bibnamefont
  {Gubser}}\ and\ \bibinfo {author} {\bibfnamefont {A.}~\bibnamefont
  {Nellore}},\ }\bibfield  {title} {\bibinfo {title} {{Mimicking the QCD
  equation of state with a dual black hole}},\ }\href
  {https://doi.org/10.1103/PhysRevD.78.086007} {\bibfield  {journal} {\bibinfo
  {journal} {Phys. Rev. D}\ }\textbf {\bibinfo {volume} {78}},\ \bibinfo
  {pages} {086007} (\bibinfo {year} {2008})},\ \Eprint
  {https://arxiv.org/abs/0804.0434} {arXiv:0804.0434 [hep-th]} \BibitemShut
  {NoStop}%
\bibitem [{\citenamefont {DeWolfe}\ \emph {et~al.}(2011)\citenamefont
  {DeWolfe}, \citenamefont {Gubser},\ and\ \citenamefont
  {Rosen}}]{DeWolfe:2010he}%
  \BibitemOpen
  \bibfield  {author} {\bibinfo {author} {\bibfnamefont {O.}~\bibnamefont
  {DeWolfe}}, \bibinfo {author} {\bibfnamefont {S.~S.}\ \bibnamefont
  {Gubser}},\ and\ \bibinfo {author} {\bibfnamefont {C.}~\bibnamefont
  {Rosen}},\ }\bibfield  {title} {\bibinfo {title} {{A holographic critical
  point}},\ }\href {https://doi.org/10.1103/PhysRevD.83.086005} {\bibfield
  {journal} {\bibinfo  {journal} {Phys. Rev. D}\ }\textbf {\bibinfo {volume}
  {83}},\ \bibinfo {pages} {086005} (\bibinfo {year} {2011})},\ \Eprint
  {https://arxiv.org/abs/1012.1864} {arXiv:1012.1864 [hep-th]} \BibitemShut
  {NoStop}%
\bibitem [{\citenamefont {Cai}\ \emph {et~al.}(2012)\citenamefont {Cai},
  \citenamefont {He},\ and\ \citenamefont {Li}}]{Cai:2012xh}%
  \BibitemOpen
  \bibfield  {author} {\bibinfo {author} {\bibfnamefont {R.-G.}\ \bibnamefont
  {Cai}}, \bibinfo {author} {\bibfnamefont {S.}~\bibnamefont {He}},\ and\
  \bibinfo {author} {\bibfnamefont {D.}~\bibnamefont {Li}},\ }\bibfield
  {title} {\bibinfo {title} {{A hQCD model and its phase diagram in
  Einstein-Maxwell-Dilaton system}},\ }\href
  {https://doi.org/10.1007/JHEP03(2012)033} {\bibfield  {journal} {\bibinfo
  {journal} {JHEP}\ }\textbf {\bibinfo {volume} {03}},\ \bibinfo {pages}
  {033}},\ \Eprint {https://arxiv.org/abs/1201.0820} {arXiv:1201.0820 [hep-th]}
  \BibitemShut {NoStop}%
\bibitem [{\citenamefont {Rougemont}\ \emph {et~al.}(2016)\citenamefont
  {Rougemont}, \citenamefont {Ficnar}, \citenamefont {Finazzo},\ and\
  \citenamefont {Noronha}}]{Rougemont:2015wca}%
  \BibitemOpen
  \bibfield  {author} {\bibinfo {author} {\bibfnamefont {R.}~\bibnamefont
  {Rougemont}}, \bibinfo {author} {\bibfnamefont {A.}~\bibnamefont {Ficnar}},
  \bibinfo {author} {\bibfnamefont {S.}~\bibnamefont {Finazzo}},\ and\ \bibinfo
  {author} {\bibfnamefont {J.}~\bibnamefont {Noronha}},\ }\bibfield  {title}
  {\bibinfo {title} {{Energy loss, equilibration, and thermodynamics of a
  baryon rich strongly coupled quark-gluon plasma}},\ }\href
  {https://doi.org/10.1007/JHEP04(2016)102} {\bibfield  {journal} {\bibinfo
  {journal} {JHEP}\ }\textbf {\bibinfo {volume} {04}},\ \bibinfo {pages}
  {102}},\ \Eprint {https://arxiv.org/abs/1507.06556} {arXiv:1507.06556
  [hep-th]} \BibitemShut {NoStop}%
\bibitem [{\citenamefont {Grefa}\ \emph {et~al.}(2021)\citenamefont {Grefa},
  \citenamefont {Noronha}, \citenamefont {Noronha-Hostler}, \citenamefont
  {Portillo}, \citenamefont {Ratti},\ and\ \citenamefont
  {Rougemont}}]{Grefa:2021qvt}%
  \BibitemOpen
  \bibfield  {author} {\bibinfo {author} {\bibfnamefont {J.}~\bibnamefont
  {Grefa}}, \bibinfo {author} {\bibfnamefont {J.}~\bibnamefont {Noronha}},
  \bibinfo {author} {\bibfnamefont {J.}~\bibnamefont {Noronha-Hostler}},
  \bibinfo {author} {\bibfnamefont {I.}~\bibnamefont {Portillo}}, \bibinfo
  {author} {\bibfnamefont {C.}~\bibnamefont {Ratti}},\ and\ \bibinfo {author}
  {\bibfnamefont {R.}~\bibnamefont {Rougemont}},\ }\bibfield  {title} {\bibinfo
  {title} {{Hot and dense quark-gluon plasma thermodynamics from holographic
  black holes}},\ }\href {https://doi.org/10.1103/PhysRevD.104.034002}
  {\bibfield  {journal} {\bibinfo  {journal} {Phys. Rev. D}\ }\textbf {\bibinfo
  {volume} {104}},\ \bibinfo {pages} {034002} (\bibinfo {year} {2021})},\
  \Eprint {https://arxiv.org/abs/2102.12042} {arXiv:2102.12042 [nucl-th]}
  \BibitemShut {NoStop}%
\bibitem [{\citenamefont {Chen}\ and\ \citenamefont
  {Huang}(2024)}]{Chen:2024ckb}%
  \BibitemOpen
  \bibfield  {author} {\bibinfo {author} {\bibfnamefont {X.}~\bibnamefont
  {Chen}}\ and\ \bibinfo {author} {\bibfnamefont {M.}~\bibnamefont {Huang}},\
  }\bibfield  {title} {\bibinfo {title} {{Machine learning holographic black
  hole from lattice QCD equation of state}},\ }\href
  {https://doi.org/10.1103/PhysRevD.109.L051902} {\bibfield  {journal}
  {\bibinfo  {journal} {Phys. Rev. D}\ }\textbf {\bibinfo {volume} {109}},\
  \bibinfo {pages} {L051902} (\bibinfo {year} {2024})},\ \Eprint
  {https://arxiv.org/abs/2401.06417} {arXiv:2401.06417 [hep-ph]} \BibitemShut
  {NoStop}%
\bibitem [{\citenamefont {Knaute}\ \emph {et~al.}(2018)\citenamefont {Knaute},
  \citenamefont {Yaresko},\ and\ \citenamefont {K{\"a}mpfer}}]{Knaute:2017opk}%
  \BibitemOpen
  \bibfield  {author} {\bibinfo {author} {\bibfnamefont {J.}~\bibnamefont
  {Knaute}}, \bibinfo {author} {\bibfnamefont {R.}~\bibnamefont {Yaresko}},\
  and\ \bibinfo {author} {\bibfnamefont {B.}~\bibnamefont {K{\"a}mpfer}},\
  }\bibfield  {title} {\bibinfo {title} {{Holographic QCD phase diagram with
  critical point from Einstein{\textendash}Maxwell-dilaton dynamics}},\ }\href
  {https://doi.org/10.1016/j.physletb.2018.01.053} {\bibfield  {journal}
  {\bibinfo  {journal} {Phys. Lett. B}\ }\textbf {\bibinfo {volume} {778}},\
  \bibinfo {pages} {419} (\bibinfo {year} {2018})},\ \Eprint
  {https://arxiv.org/abs/1702.06731} {arXiv:1702.06731 [hep-ph]} \BibitemShut
  {NoStop}%
\bibitem [{\citenamefont {Hippert}\ \emph {et~al.}(2024)\citenamefont
  {Hippert}, \citenamefont {Grefa}, \citenamefont {Manning}, \citenamefont
  {Noronha}, \citenamefont {Noronha-Hostler}, \citenamefont {Portillo~Vazquez},
  \citenamefont {Ratti}, \citenamefont {Rougemont},\ and\ \citenamefont
  {Trujillo}}]{Hippert:2023bel}%
  \BibitemOpen
  \bibfield  {author} {\bibinfo {author} {\bibfnamefont {M.}~\bibnamefont
  {Hippert}}, \bibinfo {author} {\bibfnamefont {J.}~\bibnamefont {Grefa}},
  \bibinfo {author} {\bibfnamefont {T.~A.}\ \bibnamefont {Manning}}, \bibinfo
  {author} {\bibfnamefont {J.}~\bibnamefont {Noronha}}, \bibinfo {author}
  {\bibfnamefont {J.}~\bibnamefont {Noronha-Hostler}}, \bibinfo {author}
  {\bibfnamefont {I.}~\bibnamefont {Portillo~Vazquez}}, \bibinfo {author}
  {\bibfnamefont {C.}~\bibnamefont {Ratti}}, \bibinfo {author} {\bibfnamefont
  {R.}~\bibnamefont {Rougemont}},\ and\ \bibinfo {author} {\bibfnamefont
  {M.}~\bibnamefont {Trujillo}},\ }\bibfield  {title} {\bibinfo {title}
  {{Bayesian location of the QCD critical point from a holographic
  perspective}},\ }\href {https://doi.org/10.1103/PhysRevD.110.094006}
  {\bibfield  {journal} {\bibinfo  {journal} {Phys. Rev. D}\ }\textbf {\bibinfo
  {volume} {110}},\ \bibinfo {pages} {094006} (\bibinfo {year} {2024})},\
  \Eprint {https://arxiv.org/abs/2309.00579} {arXiv:2309.00579 [nucl-th]}
  \BibitemShut {NoStop}%
\bibitem [{\citenamefont {Zhao}\ \emph {et~al.}(2024)\citenamefont {Zhao},
  \citenamefont {He}, \citenamefont {Hou}, \citenamefont {Li},\ and\
  \citenamefont {Li}}]{Zhao:2023gur}%
  \BibitemOpen
  \bibfield  {author} {\bibinfo {author} {\bibfnamefont {Y.-Q.}\ \bibnamefont
  {Zhao}}, \bibinfo {author} {\bibfnamefont {S.}~\bibnamefont {He}}, \bibinfo
  {author} {\bibfnamefont {D.}~\bibnamefont {Hou}}, \bibinfo {author}
  {\bibfnamefont {L.}~\bibnamefont {Li}},\ and\ \bibinfo {author}
  {\bibfnamefont {Z.}~\bibnamefont {Li}},\ }\bibfield  {title} {\bibinfo
  {title} {{Phase structure and critical phenomena in two-flavor QCD by
  holography}},\ }\href {https://doi.org/10.1103/PhysRevD.109.086015}
  {\bibfield  {journal} {\bibinfo  {journal} {Phys. Rev. D}\ }\textbf {\bibinfo
  {volume} {109}},\ \bibinfo {pages} {086015} (\bibinfo {year} {2024})},\
  \Eprint {https://arxiv.org/abs/2310.13432} {arXiv:2310.13432 [hep-ph]}
  \BibitemShut {NoStop}%
\bibitem [{\citenamefont {Zhu}\ \emph {et~al.}(2025)\citenamefont {Zhu},
  \citenamefont {Chen}, \citenamefont {Zhou}, \citenamefont {Zhang},\ and\
  \citenamefont {Huang}}]{Zhu:2025gxo}%
  \BibitemOpen
  \bibfield  {author} {\bibinfo {author} {\bibfnamefont {L.}~\bibnamefont
  {Zhu}}, \bibinfo {author} {\bibfnamefont {X.}~\bibnamefont {Chen}}, \bibinfo
  {author} {\bibfnamefont {K.}~\bibnamefont {Zhou}}, \bibinfo {author}
  {\bibfnamefont {H.}~\bibnamefont {Zhang}},\ and\ \bibinfo {author}
  {\bibfnamefont {M.}~\bibnamefont {Huang}},\ }\bibfield  {title} {\bibinfo
  {title} {{Bayesian inference of the critical end point in a (2+1)-flavor
  system from holographic QCD}},\ }\href {https://doi.org/10.1103/wpts-lbtr}
  {\bibfield  {journal} {\bibinfo  {journal} {Phys. Rev. D}\ }\textbf {\bibinfo
  {volume} {112}},\ \bibinfo {pages} {026019} (\bibinfo {year} {2025})},\
  \Eprint {https://arxiv.org/abs/2501.17763} {arXiv:2501.17763 [hep-ph]}
  \BibitemShut {NoStop}%
\bibitem [{\citenamefont {Li}\ and\ \citenamefont {Wang}(2025)}]{Li:2025lmp}%
  \BibitemOpen
  \bibfield  {author} {\bibinfo {author} {\bibfnamefont {Z.}~\bibnamefont
  {Li}}\ and\ \bibinfo {author} {\bibfnamefont {F.}~\bibnamefont {Wang}},\
  }\href@noop {} {\bibinfo {title} {{Phase transition of hot dense QCD Matter
  from a refined holographic EMD model}}} (\bibinfo {year} {2025}),\ \Eprint
  {https://arxiv.org/abs/2507.09113} {arXiv:2507.09113 [hep-ph]} \BibitemShut
  {NoStop}%
\bibitem [{\citenamefont {Chen}\ \emph {et~al.}(2019)\citenamefont {Chen},
  \citenamefont {Li},\ and\ \citenamefont {Huang}}]{Chen:2018vty}%
  \BibitemOpen
  \bibfield  {author} {\bibinfo {author} {\bibfnamefont {X.}~\bibnamefont
  {Chen}}, \bibinfo {author} {\bibfnamefont {D.}~\bibnamefont {Li}},\ and\
  \bibinfo {author} {\bibfnamefont {M.}~\bibnamefont {Huang}},\ }\bibfield
  {title} {\bibinfo {title} {{Criticality of QCD in a holographic QCD model
  with critical end point}},\ }\href
  {https://doi.org/10.1088/1674-1137/43/2/023105} {\bibfield  {journal}
  {\bibinfo  {journal} {Chin. Phys. C}\ }\textbf {\bibinfo {volume} {43}},\
  \bibinfo {pages} {023105} (\bibinfo {year} {2019})},\ \Eprint
  {https://arxiv.org/abs/1810.02136} {arXiv:1810.02136 [hep-ph]} \BibitemShut
  {NoStop}%
\bibitem [{\citenamefont {de~Haro}\ \emph {et~al.}(2001)\citenamefont
  {de~Haro}, \citenamefont {Solodukhin},\ and\ \citenamefont
  {Skenderis}}]{deHaro:2000vlm}%
  \BibitemOpen
  \bibfield  {author} {\bibinfo {author} {\bibfnamefont {S.}~\bibnamefont
  {de~Haro}}, \bibinfo {author} {\bibfnamefont {S.~N.}\ \bibnamefont
  {Solodukhin}},\ and\ \bibinfo {author} {\bibfnamefont {K.}~\bibnamefont
  {Skenderis}},\ }\bibfield  {title} {\bibinfo {title} {Holographic
  reconstruction of space-time and renormalization in the {AdS/CFT}
  correspondence},\ }\href {https://doi.org/10.1007/s002200100381} {\bibfield
  {journal} {\bibinfo  {journal} {Commun. Math. Phys.}\ }\textbf {\bibinfo
  {volume} {217}},\ \bibinfo {pages} {595} (\bibinfo {year} {2001})},\ \Eprint
  {https://arxiv.org/abs/hep-th/0002230} {arXiv:hep-th/0002230 [hep-th]}
  \BibitemShut {NoStop}%
\bibitem [{\citenamefont {Bazavov}\ \emph {et~al.}(2012)\citenamefont
  {Bazavov}, \citenamefont {Ding}, \citenamefont {Hegde}, \citenamefont
  {Kaczmarek}, \citenamefont {Karsch}, \citenamefont {Laermann}, \citenamefont
  {Mukherjee}, \citenamefont {Petreczky}, \citenamefont {Schmidt},
  \citenamefont {Smith}, \citenamefont {Soeldner},\ and\ \citenamefont
  {Wagner}}]{Bazavov:2012vg}%
  \BibitemOpen
  \bibfield  {author} {\bibinfo {author} {\bibfnamefont {A.}~\bibnamefont
  {Bazavov}}, \bibinfo {author} {\bibfnamefont {H.-T.}\ \bibnamefont {Ding}},
  \bibinfo {author} {\bibfnamefont {P.}~\bibnamefont {Hegde}}, \bibinfo
  {author} {\bibfnamefont {O.}~\bibnamefont {Kaczmarek}}, \bibinfo {author}
  {\bibfnamefont {F.}~\bibnamefont {Karsch}}, \bibinfo {author} {\bibfnamefont
  {E.}~\bibnamefont {Laermann}}, \bibinfo {author} {\bibfnamefont
  {S.}~\bibnamefont {Mukherjee}}, \bibinfo {author} {\bibfnamefont
  {P.}~\bibnamefont {Petreczky}}, \bibinfo {author} {\bibfnamefont
  {C.}~\bibnamefont {Schmidt}}, \bibinfo {author} {\bibfnamefont
  {D.}~\bibnamefont {Smith}}, \bibinfo {author} {\bibfnamefont
  {W.}~\bibnamefont {Soeldner}},\ and\ \bibinfo {author} {\bibfnamefont
  {M.}~\bibnamefont {Wagner}},\ }\bibfield  {title} {\bibinfo {title}
  {Freeze-out conditions in heavy ion collisions from {QCD} thermodynamics},\
  }\href {https://doi.org/10.1103/PhysRevLett.109.192302} {\bibfield  {journal}
  {\bibinfo  {journal} {Phys. Rev. Lett.}\ }\textbf {\bibinfo {volume} {109}},\
  \bibinfo {pages} {192302} (\bibinfo {year} {2012})},\ \Eprint
  {https://arxiv.org/abs/1208.1220} {arXiv:1208.1220 [hep-lat]} \BibitemShut
  {NoStop}%
\bibitem [{\citenamefont {Karthein}\ \emph {et~al.}(2026)\citenamefont
  {Karthein}, \citenamefont {Rajagopal}, \citenamefont {Pradeep}, \citenamefont
  {Stephanov},\ and\ \citenamefont {Yin}}]{Karthein:2025hvl}%
  \BibitemOpen
  \bibfield  {author} {\bibinfo {author} {\bibfnamefont {J.~M.}\ \bibnamefont
  {Karthein}}, \bibinfo {author} {\bibfnamefont {K.}~\bibnamefont {Rajagopal}},
  \bibinfo {author} {\bibfnamefont {M.~S.}\ \bibnamefont {Pradeep}}, \bibinfo
  {author} {\bibfnamefont {M.}~\bibnamefont {Stephanov}},\ and\ \bibinfo
  {author} {\bibfnamefont {Y.}~\bibnamefont {Yin}},\ }\bibfield  {title}
  {\bibinfo {title} {{Quantifying fluctuation signatures of the QCD critical
  point using maximum entropy freeze-out}},\ }\href
  {https://doi.org/10.1103/9sdb-m9xy} {\bibfield  {journal} {\bibinfo
  {journal} {Phys. Rev. D}\ }\textbf {\bibinfo {volume} {113}},\ \bibinfo
  {pages} {074010} (\bibinfo {year} {2026})},\ \Eprint
  {https://arxiv.org/abs/2508.19237} {arXiv:2508.19237 [nucl-th]} \BibitemShut
  {NoStop}%
\bibitem [{\citenamefont {Andronic}\ \emph {et~al.}(2018)\citenamefont
  {Andronic}, \citenamefont {Braun-Munzinger}, \citenamefont {Redlich},\ and\
  \citenamefont {Stachel}}]{Andronic:2017pug}%
  \BibitemOpen
  \bibfield  {author} {\bibinfo {author} {\bibfnamefont {A.}~\bibnamefont
  {Andronic}}, \bibinfo {author} {\bibfnamefont {P.}~\bibnamefont
  {Braun-Munzinger}}, \bibinfo {author} {\bibfnamefont {K.}~\bibnamefont
  {Redlich}},\ and\ \bibinfo {author} {\bibfnamefont {J.}~\bibnamefont
  {Stachel}},\ }\bibfield  {title} {\bibinfo {title} {{Decoding the phase
  structure of QCD via particle production at high energy}},\ }\href
  {https://doi.org/10.1038/s41586-018-0491-6} {\bibfield  {journal} {\bibinfo
  {journal} {Nature}\ }\textbf {\bibinfo {volume} {561}},\ \bibinfo {pages}
  {321} (\bibinfo {year} {2018})},\ \Eprint {https://arxiv.org/abs/1710.09425}
  {arXiv:1710.09425 [nucl-th]} \BibitemShut {NoStop}%
\bibitem [{\citenamefont {Nonaka}\ and\ \citenamefont
  {Asakawa}(2005)}]{Nonaka:2004pg}%
  \BibitemOpen
  \bibfield  {author} {\bibinfo {author} {\bibfnamefont {C.}~\bibnamefont
  {Nonaka}}\ and\ \bibinfo {author} {\bibfnamefont {M.}~\bibnamefont
  {Asakawa}},\ }\bibfield  {title} {\bibinfo {title} {{Hydrodynamical evolution
  near the QCD critical end point}},\ }\href
  {https://doi.org/10.1103/PhysRevC.71.044904} {\bibfield  {journal} {\bibinfo
  {journal} {Phys. Rev. C}\ }\textbf {\bibinfo {volume} {71}},\ \bibinfo
  {pages} {044904} (\bibinfo {year} {2005})},\ \Eprint
  {https://arxiv.org/abs/nucl-th/0410078} {arXiv:nucl-th/0410078} \BibitemShut
  {NoStop}%
\bibitem [{\citenamefont {Hatta}\ and\ \citenamefont
  {Ikeda}(2003)}]{Hatta:2002sj}%
  \BibitemOpen
  \bibfield  {author} {\bibinfo {author} {\bibfnamefont {Y.}~\bibnamefont
  {Hatta}}\ and\ \bibinfo {author} {\bibfnamefont {T.}~\bibnamefont {Ikeda}},\
  }\bibfield  {title} {\bibinfo {title} {Universality, the {QCD}
  critical/tricritical point and the quark number susceptibility},\ }\href
  {https://doi.org/10.1103/PhysRevD.67.014028} {\bibfield  {journal} {\bibinfo
  {journal} {Phys. Rev. D}\ }\textbf {\bibinfo {volume} {67}},\ \bibinfo
  {pages} {014028} (\bibinfo {year} {2003})},\ \Eprint
  {https://arxiv.org/abs/hep-ph/0210284} {arXiv:hep-ph/0210284} \BibitemShut
  {NoStop}%
\bibitem [{\citenamefont {Pelissetto}\ and\ \citenamefont
  {Vicari}(2002)}]{Pelissetto:2000ek}%
  \BibitemOpen
  \bibfield  {author} {\bibinfo {author} {\bibfnamefont {A.}~\bibnamefont
  {Pelissetto}}\ and\ \bibinfo {author} {\bibfnamefont {E.}~\bibnamefont
  {Vicari}},\ }\bibfield  {title} {\bibinfo {title} {Critical phenomena and
  renormalization group theory},\ }\href
  {https://doi.org/10.1016/S0370-1573(02)00219-3} {\bibfield  {journal}
  {\bibinfo  {journal} {Phys. Rept.}\ }\textbf {\bibinfo {volume} {368}},\
  \bibinfo {pages} {549} (\bibinfo {year} {2002})},\ \Eprint
  {https://arxiv.org/abs/cond-mat/0012164} {arXiv:cond-mat/0012164}
  \BibitemShut {NoStop}%
\bibitem [{\citenamefont {Wegner}(1972)}]{Wegner:1972}%
  \BibitemOpen
  \bibfield  {author} {\bibinfo {author} {\bibfnamefont {F.~J.}\ \bibnamefont
  {Wegner}},\ }\bibfield  {title} {\bibinfo {title} {Corrections to scaling
  laws},\ }\href {https://doi.org/10.1103/PhysRevB.5.4529} {\bibfield
  {journal} {\bibinfo  {journal} {Phys. Rev. B}\ }\textbf {\bibinfo {volume}
  {5}},\ \bibinfo {pages} {4529} (\bibinfo {year} {1972})}\BibitemShut
  {NoStop}%
\bibitem [{\citenamefont {Berdnikov}\ and\ \citenamefont
  {Rajagopal}(2000)}]{Berdnikov:1999slowing}%
  \BibitemOpen
  \bibfield  {author} {\bibinfo {author} {\bibfnamefont {B.}~\bibnamefont
  {Berdnikov}}\ and\ \bibinfo {author} {\bibfnamefont {K.}~\bibnamefont
  {Rajagopal}},\ }\bibfield  {title} {\bibinfo {title} {Slowing out of
  equilibrium near the {QCD} critical point},\ }\href
  {https://doi.org/10.1103/PhysRevD.61.105017} {\bibfield  {journal} {\bibinfo
  {journal} {Phys. Rev. D}\ }\textbf {\bibinfo {volume} {61}},\ \bibinfo
  {pages} {105017} (\bibinfo {year} {2000})},\ \Eprint
  {https://arxiv.org/abs/hep-ph/9912274} {arXiv:hep-ph/9912274 [hep-ph]}
  \BibitemShut {NoStop}%
\bibitem [{\citenamefont {Mukherjee}\ \emph {et~al.}(2015)\citenamefont
  {Mukherjee}, \citenamefont {Venugopalan},\ and\ \citenamefont
  {Yin}}]{Mukherjee:2015swa}%
  \BibitemOpen
  \bibfield  {author} {\bibinfo {author} {\bibfnamefont {S.}~\bibnamefont
  {Mukherjee}}, \bibinfo {author} {\bibfnamefont {R.}~\bibnamefont
  {Venugopalan}},\ and\ \bibinfo {author} {\bibfnamefont {Y.}~\bibnamefont
  {Yin}},\ }\bibfield  {title} {\bibinfo {title} {{Real time evolution of
  non-Gaussian cumulants in the QCD critical regime}},\ }\href
  {https://doi.org/10.1103/PhysRevC.92.034912} {\bibfield  {journal} {\bibinfo
  {journal} {Phys. Rev. C}\ }\textbf {\bibinfo {volume} {92}},\ \bibinfo
  {pages} {034912} (\bibinfo {year} {2015})},\ \Eprint
  {https://arxiv.org/abs/1506.00645} {arXiv:1506.00645 [hep-ph]} \BibitemShut
  {NoStop}%
\bibitem [{\citenamefont {Mukherjee}\ \emph {et~al.}(2016)\citenamefont
  {Mukherjee}, \citenamefont {Venugopalan},\ and\ \citenamefont
  {Yin}}]{Mukherjee:2016kyu}%
  \BibitemOpen
  \bibfield  {author} {\bibinfo {author} {\bibfnamefont {S.}~\bibnamefont
  {Mukherjee}}, \bibinfo {author} {\bibfnamefont {R.}~\bibnamefont
  {Venugopalan}},\ and\ \bibinfo {author} {\bibfnamefont {Y.}~\bibnamefont
  {Yin}},\ }\bibfield  {title} {\bibinfo {title} {{Universal off-equilibrium
  scaling of critical cumulants in the QCD phase diagram}},\ }\href
  {https://doi.org/10.1103/PhysRevLett.117.222301} {\bibfield  {journal}
  {\bibinfo  {journal} {Phys. Rev. Lett.}\ }\textbf {\bibinfo {volume} {117}},\
  \bibinfo {pages} {222301} (\bibinfo {year} {2016})},\ \Eprint
  {https://arxiv.org/abs/1605.09341} {arXiv:1605.09341 [hep-ph]} \BibitemShut
  {NoStop}%
\end{thebibliography}%

\end{document}